\documentclass[aps,pra,reprint,groupedaddress,nobibnotes,nofootinbib,longbibliography]{revtex4-1}
\usepackage[english]{babel}
\usepackage[utf8x]{inputenc}
\usepackage[T1]{fontenc}
\usepackage{url}
\usepackage{graphicx}
\graphicspath{{figures/}}
\usepackage{amsmath} 
\usepackage{amssymb} 
\usepackage[detect-all]{siunitx}
\usepackage{bm}
\usepackage{mathtools}
\usepackage{color}
\usepackage{layouts}
\renewcommand{\d}{\mathrm{d}}
\usepackage{amsthm}

\usepackage{fancyhdr}
\begin{document}
	
\title{Walker breakdown with a twist: Dynamics of multilayer domain walls and skyrmions driven by spin-orbit torque.}
\author{Ivan Lemesh}
\email{ivan.g.lemesh@gmail.com}
\affiliation{Department of Materials Science and Engineering, Massachusetts Institute of Technology, Cambridge, Massachusetts 02139, USA}
\author{Geoffrey S. D. Beach}
\affiliation{Department of Materials Science and Engineering, Massachusetts Institute of Technology, Cambridge, Massachusetts 02139, USA}
\date{\today}
\begin{abstract}
Current-induced dynamics of twisted domain walls and skyrmions in ferromagnetic perpendicularly magnetized multilayers is studied through three-dimensional micromagnetic simulations and analytical modeling. It is shown that such systems generally exhibit a Walker breakdown-like phenomenon in the presence of current-induced damping-like spin-orbit torque. Above a critical current threshold, corresponding to typical velocities of the order tens of m/s, domain walls in some layers start to precess with frequencies in the gigahertz regime, which leads to oscillatory motion and a significant drop in mobility. This phenomenon originates from complex stray field interactions and occurs for a wide range of multilayer materials and structures that include at least three ferromagnetic layers and finite Dzyaloshinskii-Moriya interaction. An analytical model is developed to describe the precessional dynamics in multilayers with surface-volume stray field interactions, yielding qualitative agreement with micromagnetic simulations. 
\end{abstract}
\maketitle 

\section{Introduction}
Spintronic devices may provide a path to achieving high data density, ultralow power consumption, and high-speed operation in beyond-CMOS data storage and computing technologies~\cite{Hirohata2014}. Magnetic domain walls (DWs)~\cite{Neel1955,Koyama2011} and skyrmions~\cite{Belavin1975,Nagaosa2013}, localized twists of the magnetization with particle-like characteristics, are of high interest as potential information carriers in spintronic devices, owing to their topological properties and facile manipulation by electric currents.  In particular, the small size, enhanced stability, and ability to follow two-dimensional trajectories make skyrmions extremely promising for racetrack storage~\cite{Parkin2008,Fert2013,Sampaio2013,Tomasello2014,Wiesendanger2016} or novel non-von Neumann computing architectures~\cite{Zazvorka2019,Pinna2018,Bourianoff2018,Prychynenko2018,Kyung2019}.  Pioneering early work on magnetic skyrmions focused on bulk noncentrosymmetric materials~\cite{Jeong2004,Uchida2006,Yu2010} with low ordering temperatures, or ultrathin metal films in which nanoscale skyrmions can be stabilized at a low temperature~\cite{Heinze2011}.  Recently, it has been found that multilayers with perpendicular magnetic anisotropy (PMA) can host magnetic skyrmions at room termperature~\cite{Jiang2015,Woo2016,Moreau-Luchaire2016,Boulle2016}. Enhanced skyrmion stability in multilayer films is afforded by the increased skyrmion volume when the total film thickness is increased~\cite{Buttner2018}. Interfaces in such films give rise to the Dzyaloshinskii-Moriya interaction (DMI), which promotes the N\'eel character of spin textures~\cite{Haazen2013,Emori2013,Ryu2013,Yang2015}, and to a dampinglike spin-orbit torque (SOT), which provides for their efficient current-driven motion.

Although the static behaviors of multilayer skyrmions are now reasonably well-understood, their dynamics has been less studied, despite the critical role that the dynamics plays in terms of potential applications.  In single ferromagnet/heavy-metal bilayers, DWs driven by SOT tend to maintain dynamic equilibrium, i.e., the DW plane is characterized by a fixed (but current-dependent~\cite{Boulle2013}) angle $\psi$. This is in sharp contrast with the phenomenon of Walker breakdown (DW precession) that occurs in bubbles and straight DWs driven by magnetic fields~\cite{Schryer1974,Malozemoff1979,Beach2005} or spin-transfer torques (STT)~\cite{Berger1978,Zhang2004,Thiaville2004,Mougin2007} that exceed a critical threshold. Walker breakdown is precluded by symmetry for SOT-driven motion in conventional single ultrathin ferromagnetic layers~\cite{Linder2013,Risinggard2017}, since at high drive, $\psi$ tends asymptotically toward the hard axis but is never driven into precession. However, recently it has been found that in in multilayers of ferromagnet and heavy-metal, DWs and skyrmions can exhibit through-thickness twists~\cite{Dovzhenko2016,Montoya2017,Legrand2017a,Lemesh2018,Legrand2018,Montoya2018} such that the statics and dynamics can no longer be described using a single value of $\psi$.  Micromagnetic simulations of such twisted multilayer skyrmions~\cite{Lemesh2018} have evidenced dynamical instabilities reminiscent of Walker breakdown during SOT-driven motion, wherein Bloch line nucleation and motion in a subset of layers leads to a significantly diminished skyrmion velocity and skyrmion Hall angle. These behaviors are a result of complex surface-volume stray field interactions whose influence on the dynamics remains largely unexplored. 

In this work, we show that DWs in asymmetrically stacked ferromagnetic multilayers with PMA, in contrast to single-layer and bilayer thin films~\cite{Linder2013,Risinggard2017}, generally exhibit a Walker-breakdown-like phenomenon even when driven solely by dampinglike SOT. This breakdown occurs when certain (Bloch-like) layers reach a critical velocity, beyond which precession sets in, leading to an oscillatory trajectory and a diminished mobility.  For typical material parameters, this velocity is of the order of tens of m/s, corresponding to the velocity range in recent experiments~\cite{Woo2016,Litzius2016,Legrand2017,Hrabec2017,Woo2017}. The breakdown originates from the interplay of SOT, DMI, and magnetostatic interactions~\cite{Lemesh2018}, thanks to which DWs in some layers can be driven toward a Bloch configuration amenable to precession (because of the surface-volume stray field interactions~\cite{Lemesh2018}) while others maintain a N\'eel (or transient~\cite{Lemesh2017}) character.  This, in turn, allows the dampinglike SOT, which acts as an effective field $\propto\sin(\psi)$, to continue to drive the magnetostatically-coupled composite structure even though the Bloch layers are not directly susceptible to the driving torque. These results hence identify a critical deficiency in proposals to utilize magnetostatically-coupled multilayers for room-temperature skyrmion-based devices, thus, providing a materials engineering framework for maximizing the dynamical stability of skyrmions. 

Here, we present three-dimensional (3D) micromagnetic simulations of the current-driven dynamics of multilayer DWs and skyrmions and develop an analytical model that describes the key features. The DW velocity and precession onset predicted by our model are in good qualitative agreement with full three-dimensional micromagnetic simulations of DW and skyrmion dynamics. We hence provide essential analytical insight and predictive capability that allow for a mechanistic understanding of these newly discovered complex dynamical phenomena. Our results have important implications for the potential use of multilayer-based skyrmions in racetrack devices in which high-speed motion is desired and provide a framework for designing the dynamics of multilayer skyrmions to enable optimal behaviors.
\begin{figure*}
	\includegraphics[width=1.0\linewidth]{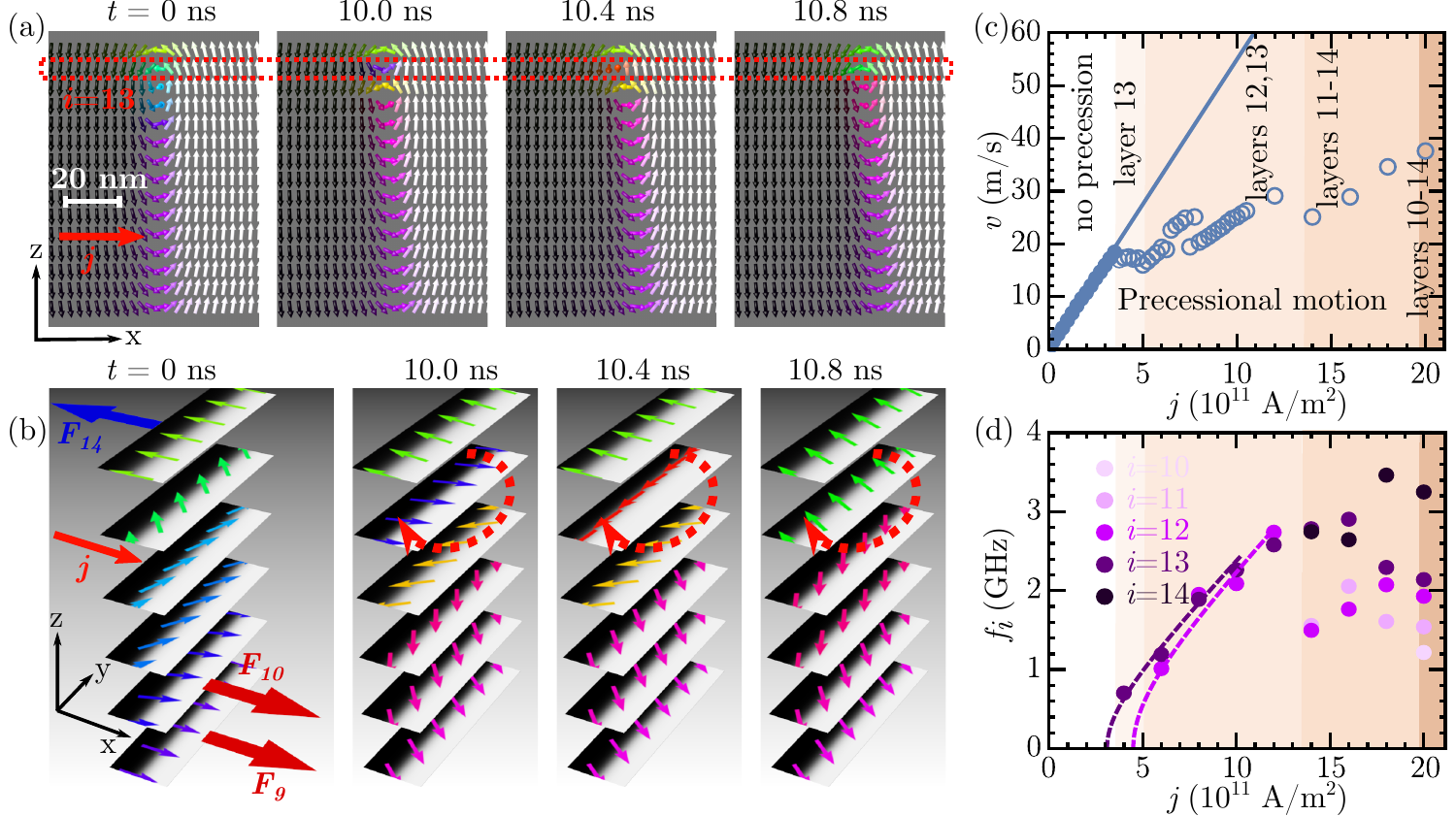}
	\caption{{\bf Time evolution of current-driven isolated twisted domain walls.} {\bf(a, b)} Micromagnetic snapshots that depict precessional motion for $j=\SI{4.5e11}{A/m^2}$ (just above the critical current). {\bf(a)} The $x$-$z$-cross section and {\bf(b)} schematic side view of transient layers.  Time $t=0$ corresponds to the static profile and the precession initiates in layer $i=13$ (arrows depict the direction of DW rotation). In the layer below it ($i=12$), the DW oscillates (without precession), while in all other layers, the DW profile remains unchanged (although quick readjustments of $\psi_i$ are still present as discussed at the end of section~\ref{sec:five}). Flat arrows show schematically the direction of the Thiele effective force $F_{i}$ from the damping-like spin-orbit torque acting on several layers.  {\bf(c)} Average DW velocity as a function of current density, with empty (filled) lines or dots indicating the presence (absence) of full $2\pi$ precession and the solid line corresponding to the steady-state DW solution~\cite{Lemesh2018}. {\bf(d)} Precessional frequency as a function of current density and layer number with dashed lines representing guides ($f\sim \sqrt{j^2-j_{cr}^2}$) to the eye. Interfacial DMI is fixed to $D=\SI{1.0}{mJ/m^2}$.  The shading in panels (c) and (d) indicate which layers, if any, are precessing, as labeled in panel (c).}\label{fig:1}
\end{figure*}

\section{Methods}
Micromagnetic simulations are performed using the Mumax3~\cite{Vansteenkiste2014} software. Material parameters are $\mathcal{T}=\SI{1}{nm}$, $M_s=\SI{1.4e6}{A/m}$, $A=\SI{1.0e-11}{J/m}$. For skyrmion dynamics simulations, the modified Slonczewski-like torque module has been used (with enabled damping-like torque corresponding to $\theta_{SH}=0.1$ and the fixed layer polarization along the −y direction). Unless specified otherwise, $\mathcal{P}=\SI{6}{nm}$ ($f=1/6$), $K_u=\SI{1.72e6}{J/m^3}$ ($Q=1.4$), $\alpha=0.3$. 

For skyrmions (domain walls) the cell size is $1.5~\text{nm}\times 1.5~\text{nm} \times 1~\text{nm}$ ($1.5~\text{nm}\times 15~\text{nm} \times 1~\text{nm}$) with the simulation size of $1125~\text{nm}\times 1125~\text{nm}\times \mathcal{N}\mathcal{P}$ and periodic boundary conditions applied in the $x$- and $y$- directions. We note that the large cell size in the y direction for the DW simulations prevents Bloch line nucleation, leading to uniform precession that is more readily treated analytically.  When using smaller ($1.5~\text{nm}$) cells in such simulations, but maintaining periodic boundary conditions, Bloch line formation is much more random than in the skyrmion simulations, sometimes occurring and sometimes not, which inhibits the systematic analysis of the dynamics.  This observation is attributed to the symmetry and continuity of the DW simulations (with periodic boundary conditions), which artificially inhibits the formation of Bloch lines.  We note that when omitting periodic boundary conditions in the y direction and using a reduced width to simulate a racetrack, the 3D simulations tend to become unstable as the DW decoupling from layer to layer is more pronounced.  This problem does not occur in the full micromangetic skyrmion simulations, as skyrmions tend to be more rigid.  

The differential equations in the analytical model are solved numerically using the explicit NDSolve method of the Wolfram Mathematica 11.3 software.

\section{Micromagnetic simulations}
We first consider the dynamics of an isolated straight DW in a multilayer film with ultrathin magnetic (M) layers ($\mathcal{T}<l_{ex}$) of thickness $\mathcal{T}=1~\text{nm}$, consisting of $\mathcal{N}=15$ multilayer repeats with a period of $\mathcal{P}=6~\text{nm}$ separated by nonmagnetic spacer layers. Although the composition of spacer layers has no effect on the DW analysis, here we imply that they consist of heavy metal layers (H) and symmetry breaking layers (S) incorporated into an asymmetrically stacked heterostructure of [H/M/S]$_{\mathcal{N}}$-type, similar to those studied in a number of recent experimental works in which room-temperature skyrmions have been stabilized~\cite{Woo2016,Moreau-Luchaire2016,Litzius2016,Legrand2017,Legrand2017a,Buttner2017e,Lemesh2018a,Kyung2019}. We assume a saturation magnetization $M_s=\SI{1.4e6}{A/m}$, quality factor $Q=2K_u/\mu_0M_s^2=1.4$ (where $K_u$ is the uniaxial magnetocrystalline anisotropy constant and $\mu_0$ is the vacuum permeability), exchange stiffness $A=\SI{1.0e-11}{J/m}$, and interfacial DMI, $D=\SI{1.0}{mJ/m^2}$, representative of typical experimental skyrmion-hosting multilayers~\cite{Buttner2015b,Woo2016,Moreau-Luchaire2016a,Litzius2016,Wiesendanger2016,Buttner2017e}.  The static DW profile in such a material exhibits a twisted character as shown elsewhere~\cite{Dovzhenko2016,Legrand2017a,Lemesh2018} and depicted in Fig.~\ref{fig:1}(a).  The DW profile varies from N\'eel of one chirality ($\sin(\psi)=1$) at layer number $i=0$ to N\'eel of the opposite chirality ($\sin(\psi)=-1$) at $i=14$, with layers 12 and 13 exhibiting a Bloch-like character ($\sin(\psi)\approx 0$).

We perform full 3D micromagnetic simulations (see Methods) of current-driven motion by damping-like SOT, assuming a spin Hall angle $\theta_{\text{SH}}=0.1$~\cite{Emori2013,Liu2012a} and damping constant $\alpha=0.3$~\cite{Yuan2003,Metaxas2007,Schellekens2013}.  Here, the SOT is provided by the charge current $j$ that flows in the heavy metal layer along the $x$-direction (see Fig.~\ref{fig:1}(a)). The simulations reveal that for small current densities ($j<\SI{3.75e11}{A/m^2}$), the DW translates uniformly with a linear trajectory (position versus time) at a velocity proportional to $j$ (Fig.~\ref{fig:1}(c)). The DW angles ($\psi_i$) slightly readjust in all layers in accordance with the new dynamic equilibrium, but the profile of the twist remains qualitatively the same. The situation changes drastically once the current exceeds a threshold $j_{cr}\approx\SI{3.75e11}{A/m^2}$, at which point the DW at layer $i=13$ begins to precess (Figs.~\ref{fig:1}(a),(b)). The frequency of this (nonuniform) precession is of the order of approximately 1~GHz  as shown in Fig.~\ref{fig:1}(d). 

\begin{figure}
	\includegraphics[width=1.0\linewidth]{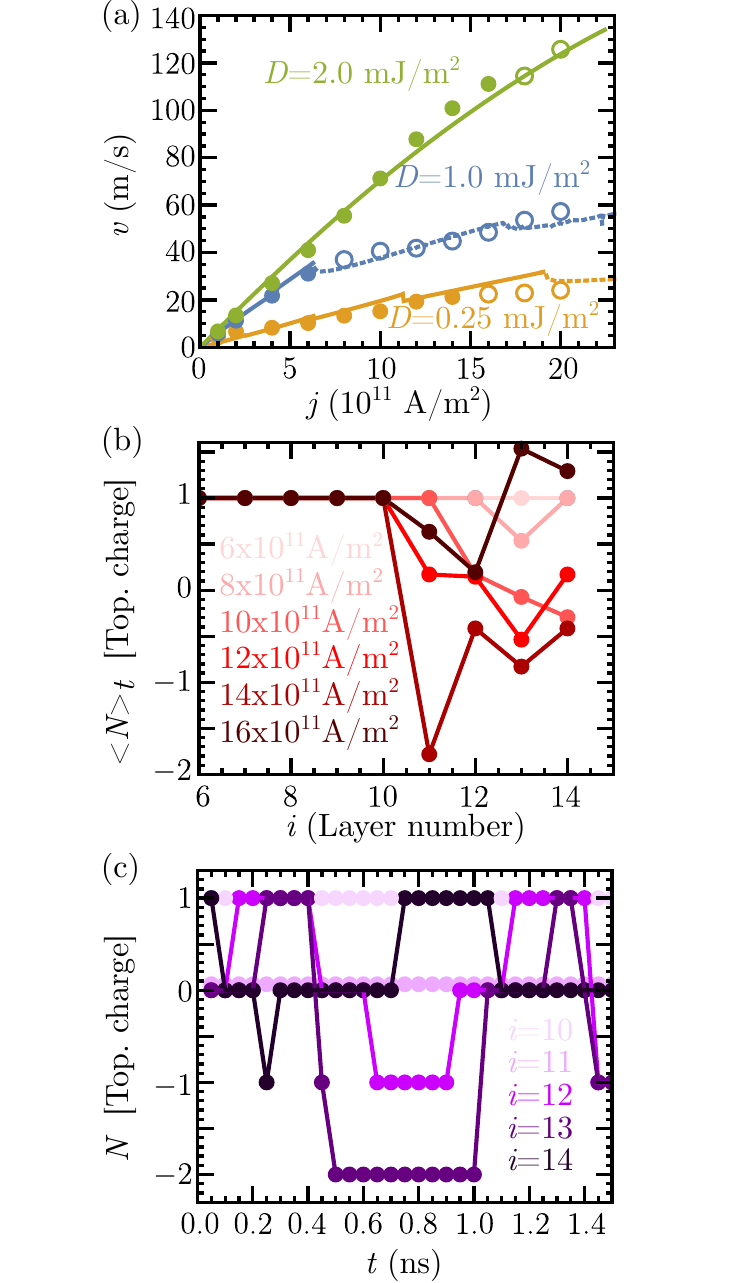}
	\caption{{\bf The dynamic properties of twisted skyrmions.} {\bf(a)} Skyrmion velocity as a function of current density. Continuous lines correspond to analytical DW model calculations, and symbols correspond to 3D micromagnetic simulations for skyrmions, with closed symbols indicating precession-free motion and open symbols denoting precessional motion. {\bf(b)}  Average topological charge as a function of layer number and current density for $D=\SI{1.0}{mJ/m^2}$. The topological charge is computed as 	$N = 1/4\pi\int \mathbf{m}\cdot\left(\partial \mathbf{m}/\partial x\times \partial \mathbf{m}/\partial y \right) \d x \d y $~\cite{Belavin1975}. {\bf(c)} Instantaneous topological charge as a function of time for $D=\SI{1.0}{mJ/m^2}$ and $j=\SI{12e11}{A/m^2}$. }\label{fig:3}
\end{figure}

Above $j_{cr}$, the DW translates with an oscillatory trajectory.  The corresponding average velocity $v(j)$ becomes nonlinear in this regime, with a slight drop at $j_{cr}$ followed by a (sublinear) increase at higher $j$ (Figs.~\ref{fig:1}(c)). The precessional frequency of the precessing layer monotonically increases with increasing $j$, as indicated in Fig.~\ref{fig:1}(d). At higher currents, more layers begin to precess, resulting in additional Walker breakdown-like features in $v(j)$. For instance, at $j\approx\SI{5e11}{A/m^2}$, the precession also initiates in layer $i=12$.  This behavior leads to a substantial reduction in the velocity compared to that expected from extrapolation from the low-$j$ mobility (slope of $v$ vs $j$) in the absence of precession.  
\begin{figure*}
	\includegraphics[width=1.0\linewidth]{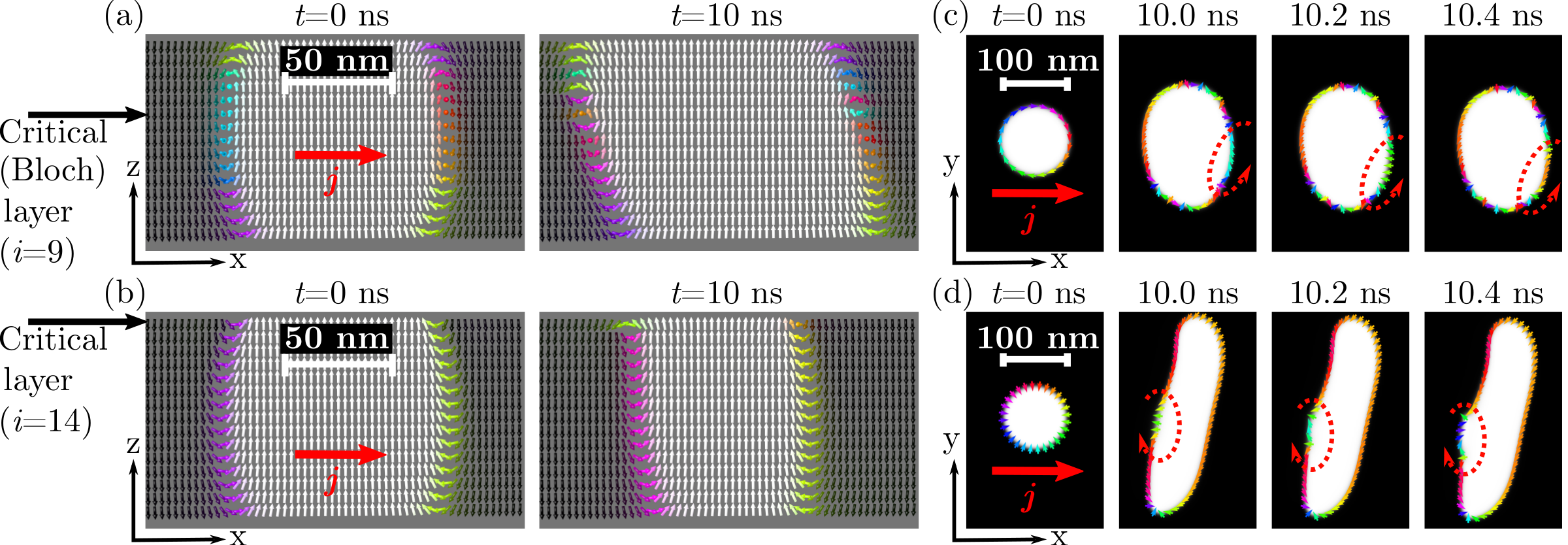}
	\caption{{\bf The evolution of twisted skyrmions right above the critical current.} {\bf(a, b)} $x$-$z$ cross sections of low- and high DMI skyrmions and {\bf(c, d)} in-plane micromagnetic profiles of the precessing layer (arrows depict the direction of spin rotation). The upper (lower) row corresponds to $D=\SI{0.25}{mJ/m^2}$, $j=\SI{12.5e11}{A/m^2}$, $i=9$ ($D=\SI{2.0}{mJ/m^2}$, $j=\SI{17.0e11}{A/m^2}$, $i=14$). The corresponding DW width, ($\Delta_i$) and DW angle ($\psi_i$) are depicted as a function of DMI in Fig.~1 of Ref.~\cite{Lemesh2018}.}
	\label{fig:2}
\end{figure*}
Similar behavior is observed for stray field skyrmions~\cite{Buttner2018}, as seen in the micromagnetic snapshots in Fig.~\ref{fig:2} and the $v(j)$ curves in Fig.~\ref{fig:3}(a).  Here, we performed 3D micromagnetic simulations using the same material parameters as above, except for the DMI, which is varied in the range $D=0$ to $D=\SI{3.0}{mJ/m^2}$. For each DMI value, the magnetic field was adjusted to yield the same equilibrium skyrmion radius of $R\approx50~\text{nm}$. At $D=0$ the Bloch layer is in the center of the film and since the SOT on the top half of the film cancels that in the bottom half, the skyrmion is immobile.  Increasing $D$ shifts the position of the Bloch layer upward, as shown previously~\cite{Lemesh2018}, leading to a nonzero net Thiele effective force~\cite{Thiele1973} from the SOT to drive the skyrmion.  We find that Walker breakdown-like behavior occurs generally in multilayers with $D>0$, where it is mediated by the generation and motion of Bloch lines in the DW that bounds the bubblelike skyrmion.  

Figure~\ref{fig:2} shows micromagnetic snapshots during SOT-driven motion just above $j_{cr}$ for the cases $D=0.25$ and $D=\SI{2.0}{mJ/m^2}$.  These simulations correspond, respectively, to cases where a Bloch layer exists near the center of the film (low DMI) and where the entire stack is N\'eel (high DMI).  Corresponding $v(j)$ curves are shown in Fig.~\ref{fig:3}(a).  Precessional motion tends to initiate in either the Bloch-like layer for intermediate DMI or in the top-most layer once the DMI exceeds the threshold at which all other layers are N\'eel.  The critical current tends to increase for larger $D$, as seen in Fig.~\ref{fig:3}(a)). 
We find that current also leads to an elongation of the skyrmion, which increases as the current density increases. This accounts for the larger distortion seen in Fig.~\ref{fig:2}(d) compared to Fig.~\ref{fig:2}(c), where a larger driving current is used in the latter simulation to drive the system into precession. The corresponding eccentricity of a skyrmion as a function of current and DMI are visualized as contours plotted in Figure~\ref{fig:6}(a). These distorted objects maintain dynamic equilibrium during the injection of current. However, when the current is switched off, they go back to their original circular shape, with a diameter set by the applied field.
	
From Fig.~\ref{fig:3}(a), we can also observe that the velocities of simulated skyrmions are limited to the same $v_{cr}$ as given by the analytical DW theory (as discussed next). This indicates that precessing DWs play a defining velocity-limiting role in skyrmion propagation. Indeed, such precessing DWs can be found in two radial sections of every supercritical skyrmion. Similarly to our observations for multilayer DWs, as $j$ increases past $j_{cr}$, additional layers begin to precess, in this case through the creation of Bloch lines\footnote{Micromagnetic tools can provide only a qualitative understanding of the process of Bloch line/point nucleation. For quantitative analysis, atomistic simulations should be used.} in additional layers.  This implies that the topological charge for skyrmions beyond this threshold is not fixed, but rather varies with time.  This is seen in Figs.~\ref{fig:3}(b) and (c) which show the time-averaged topological charge in each layer at various current densities (Fig.~\ref{fig:3}(b)) and the dynamical evolution of the topological charge ((Fig.~\ref{fig:3}(c)) for an exemplary case $D=\SI{1}{mJ/m^2}$.  Hence, our results imply that multilayer-based skyrmions may not be topologically well-defined dynamically, even when driven at relatively low velocities. 

\section{Analytical model for multilayer domain walls}
The observed behaviors have all the hallmarks of Walker breakdown, a well-known transition from steady-state to precessional dynamics exhibited by DWs once they reach a threshold velocity $v_{cr}$~\cite{Schryer1974,Malozemoff1979,Thiaville2012}. Walker breakdown is however not expected for damping-like SOT-driven motion, at least in 2D systems.  This threshold is related to the ``stiffness’’ of $\psi$ against rotation away from its equilibrium orientation, which can be characterized by an effective field $H_{\text{stiff}}$ acting on the DW moment.  In the one-dimensional DW model \cite{Schryer1974,Malozemoff1979}, $v_{cr}=\gamma \mu_0 \Delta H_{\text{stiff}}/2$, where in the case of strong DMI, $H_{\text{stiff}}$ is approximately equal to twice the DMI effective field, $2H_D$ \cite{Thiaville2012}.  Hence, the Walker threshold cannot be reached for SOT-driven DWs in single-layer films, since $\gamma \mu_0 \Delta H_D$ also corresponds to the asymptotic velocity limit imposed by the dampinglike SOT symmetry~\cite{Thiaville2012} (see Eq.~\eqref{eq:singlesdmi} in Appendix~\ref{sec:walker}). 

In multilayers, however, we see that the Walker limit can indeed be reached.  The process, which we treat in detail below, can be understood qualitatively as follows.  We find that precession initiates in the layer at which the sum of the interfacial DMI ($D$) and the DMI-like surface-volume stray field energy component ($D_{sv}$~\cite{Lemesh2018}),  
\begin{align}
	D_{\text{eff}}(i) &=D+D_{sv}(i)\label{eq:thr1},\\
	D_{sv} (i)&=-\frac{\mathcal{N}}{\pi  f}\sum_{j=0}^{\mathcal{N}-1}  F_{sv, ij}(\Delta)\text{sgn}(i-j)\label{eq:thr2}.  
\end{align} 
is minimum. This corresponds to the most Bloch-like layer ($i_{cr}=i_{\text{Bloch}}$) in the case of a twisted DW structure, or to the topmost layer ($i_{cr}=(\mathcal{N}-1)\delta_{1, \text{sgn}(D)}$) when all other layers are N\'eel.

Since $D_{\text{eff}}(i)$ is reduced from $D$, and is close to zero for certain layers in twisted DWs, the effective field that supplies the restoring torque on a driven DW in those layers is small.  Although the SOT cannot directly drive such Bloch layers due to its symmetry, in the case of a multilayer, as depicted in Fig.~\ref{fig:1}(b), there is a net Thiele effective force due to the magnetostatic coupling and the dampinglike SOT acting on the other layers.  Thus, as long as there is finite DMI that ensures there are more N\'eel-like DWs of one orientation than the other, then the least ``stiff'' layers can be driven beyond $v_{cr}$.

To model this behavior analytically, we treat the DW as a classical object with the Lagrangian density $\mathcal{L}$ and the Rayleigh dissipation functional $\mathcal{F}$ expressed similarly to the single-layer case~\cite{Boulle2011,Boulle2013}, 
\begin{align}
	\mathcal{L} & =  \sigma_{tot}^{ 1, \mathcal{N}} - \frac{M_sf}{\gamma\mathcal{N}}\sum_{i=0}^{\mathcal{N}-1} \int_{-\infty}^{+\infty} \d x\left\{\left(\psi_i-\pi/2\right) \sin(\theta_i)\frac{{\d \theta_i}}{\d t} \right\},\label{eq:lagrangian}\\
	\mathcal{F} & = \frac{\alpha M_sf}{2\gamma \mathcal{N}}\sum_{i=0}^{\mathcal{N}-1}\int_{-\infty}^{+\infty} \d x\left[\frac{\d\mathbf{m_i}}{\d t}-\frac{\gamma}{\alpha}B_{\text{DL}}\mathbf{m_i}\times \mathbf{\hat{y}}\right]^2\label{eq:diss}.
\end{align}
where $\gamma=\SI{1.76e11}{sA/kg}$ is the gyromagnetic ratio, $f=\mathcal{T}/\mathcal{P}$ is the scaling factor, $B_{\text{DL}}=\frac{\hbar\theta_{\text{SH}}}{2eM_s \mathcal{T}}j$ is the damping like SOT effective field, $\hbar$ is the reduced Planck constant, and $\sigma_{tot}^{ 1, \mathcal{N}}  = \int_{-\infty}^{+\infty}\mathcal{E}_{tot}^{ 1, \mathcal{N}}(x)\d x $ is the total cross-sectional energy density of the DW normalized per single layer repeat. In Appendix~\ref{sec:derivations}, we consider a more general equation in the presence of STT and magnetic fields, while here we focus on the specific case of a field-free system with an in-plane current that carries only the dampinglike SOT.
\begin{figure}
	\includegraphics[width=1.0\linewidth]{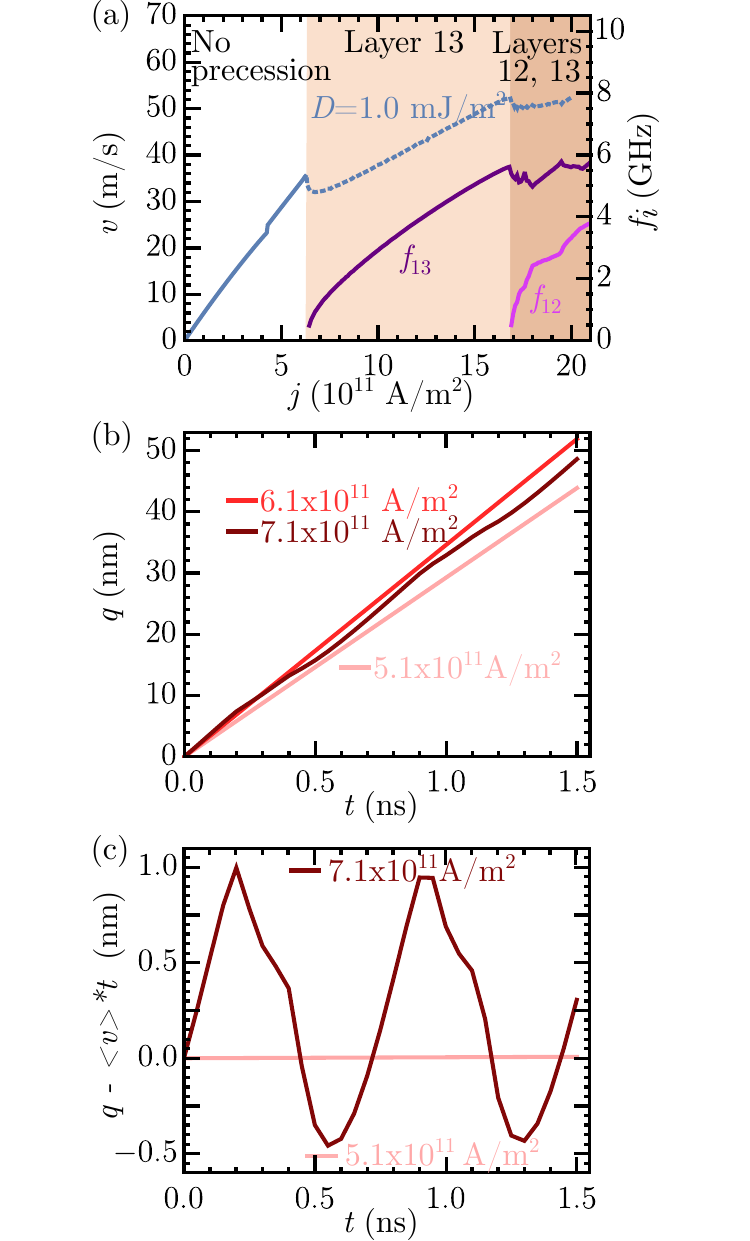}
	\caption{
			{\bf The dynamics of twisted DWs given by analytical theory.} {\bf(a)} DW velocity and precessional frequency as a function of current density. {\bf(b)} DW position as a function of time and current density in the proximity of the critical current ($j_{cr}=\SI{6.4e11}{A/m^2}$). {\bf(c)} The same as in {\bf(b)} but after subtracting the linear $\langle v \rangle * t$ contribution, where $\langle v \rangle$ is the time-averaged velocity and $t$ is the time. For all cases, $D=\SI{1.0}{mJ/m^2}$. }\
		\label{fig:4}
\end{figure}
We assume that the DW in every layer follows a Walker profile~\cite{Schryer1974}, with a constant DW angle $\psi_i$ and a polar angle $\theta_i(x) = \arctan\{\exp[\mp(x-q_i)/\Delta_i]\} $, where upper (lower) sign stands for $\downarrow|\uparrow$ ($\uparrow|\downarrow)$ DW state. Assuming that all the layers are perfectly coupled ($q_i\equiv q$) and have identical width ($\Delta_i\equiv\Delta$), we can use the result of our earlier work~\cite{Lemesh2018}, where we showed that in magnetic multilayers, the total energy of twisted straight DWs can be expressed as 
\begin{align} \sigma_{tot}^{ 1, \mathcal{N}} (\Delta,& \psi_i)=\frac{2A}{\Delta} f +2K_{u} \Delta f \mp  \frac{\pi D f}{\mathcal{N}}\sum_{i = 0}^{\mathcal{N}-1}  \sin(\psi_i)\notag\\
	&+ \sum_{i = 0}^{\mathcal{N}-1} \sum_{j = 0}^{\mathcal{N}-1} \left\{F_{s, ij} (\Delta) +\sin(\psi_i)\sin(\psi_{j})  F_{v, ij} (\Delta)\notag \right.\\&\left.\pm \sin(\psi_i)\text{sgn}(i-j)F_{sv, ij} (\Delta)\right\} \label{eq:exactstray}
\end{align}
with generic functions $F_{\alpha, ij}$ defined in Ref.~\cite{Lemesh2018} (and Eq.~\eqref{eq:A2} in Appendix~\ref{sec:derivations}). Using the Walker profile, we can also integrate Eqs.~\eqref{eq:lagrangian},~\eqref{eq:diss}, which results in analytical expressions for $\mathcal{L}(\Delta, \psi_i)$ and $\mathcal{F}(\Delta, \psi_i) $ as shown in Appendix~\ref{sec:derivations}. The equation of motion can then be obtained from the Lagrange-Rayleigh equations using the generalized coordinates $X=q, \psi_i, \Delta$:
\begin{align}
	\frac{\partial\mathcal{L}}{\partial X}-\frac{\partial}{\partial t}\frac{\partial\mathcal{L}}{\partial \dot{X}}-\frac{\partial}{\partial x}\frac{\partial\mathcal{L}}{\partial X'}+\frac{\partial\mathcal{F}}{\partial \dot{X}}&=0,
\end{align}
Considering the case of a {\it freely propagating} DW ($\dot{q}\neq0$) with time independent width ($\dot{\Delta}=0$), these equations reduce to
\begin{align}
	\dot{q}&= \frac{\mp \Delta}{\alpha \mathcal{N}\cos(\chi)} \sum_{j=0}^{\mathcal{N}-1}\left[\frac{\pi \gamma B_{\text{DL}} }{2}  \sin(\psi_j)- \dot{\psi_j}\right]\label{eq:qdyn3} 
\end{align}
where $\psi_i(t)$ can be found from the following $\mathcal{N}$ equations:
\begin{align}
	&\sum_{j=0}^{\mathcal{N}-1} \left[\dot{\psi_j}\left( \frac{1}{\mathcal{N}} + \alpha^2\delta_{ij}\right) -\sin(\psi_j) \frac{\pi \gamma B_{\text{DL}}}{2\mathcal{N}} \right]\notag \\ &=  \frac{\alpha \gamma \cos(\psi_i-\chi)}{2M_s\Delta}  \left(\pm \pi D- \frac{\mathcal{N}}{f}\times \notag \right.\\  &  \left. \times \sum_{j=0}^{\mathcal{N}-1}\left[(1+\delta_{ij})\sin(\psi_{j}-\chi)  F_{v, ij}  \pm  \text{sgn}(i-j)F_{sv, ij}\right]\right)\label{eq:psidyn3}.
\end{align}
Here, $\Delta$ is the average DW width (described by Eq.~\eqref{eq:deltadyn0} in Appendix~\ref{sec:derivations}), which can be approximated from static equations (see Eqs.~(5, 6) in Ref.~\cite{Lemesh2018}), since it depends only weakly on $j$. Note that we also introduced the DW tilt $\chi$~\cite{Boulle2013}, although in contrast to Ref.~\cite{Boulle2013}, $\chi$ here is a fixed parameter rather than a conjugate variable. We find from numerics that the critical current (when present) takes a minimum value at $\chi=0$, which corresponds to the straight transverse DW. This is in line with the fact that for a skyrmion, the precession typically initiates nearby its front and back edges, as seen in Fig.~\ref{fig:2}. Hence, for our further analysis, we focus on the case $\chi=0$.

The steady state analysis of Eqs.~\eqref{eq:qdyn3},~\eqref{eq:psidyn3} predicts that the Walker breakdown phenomenon is generally present in films with $\mathcal{N}>2 $ and is absent for single or bi-layers (as shown in Appendix~\ref{sec:walker}). The resulting numerical solution of Eqs.~\eqref{eq:qdyn3},~\eqref{eq:psidyn3} is depicted in Fig.~\ref{fig:4}, using the same material parameters as those used for the simulations in Fig.~\ref{fig:1}. Our theoretical model accurately captures the critical layer number in which precession originates, although at slightly higher current densities compared with micromagnetic simulations. It also captures the monotonic increase of the precession frequency with current.  Above the precessional threshold, we see a transition from stationary translational motion to oscillatory motion (Figs.~\ref{fig:4}(b), (c)), as occurs in conventional Walker breakdown, and is evidenced in our micromagnetic simulations.  We note that at higher currents, micromagnetic simulations generally result in a larger number of precessing layers than predicted by our model, which we attribute to three factors: (i) at high currents, DWs tend to decouple laterally, as is evident from Fig.~\ref{fig:1}(a) and Fig.~\ref{fig:2} (also see Figs.~\ref{fig:7}(b),(c) in Appendix~\ref{sec:extra}), which fundamentally affects their dynamics,  (ii) in simulations, the cell size is finite, and (iii) our analytical equations are generally more constrained compared with micromagnetic simulations.

\section{Onset of precessional dynamics}~\label{sec:five}
Our model, though developed for straight DWs, also accurately predicts the $v(j)$ characteristics for skyrmions, as seen in Fig.~\ref{fig:3}(a), where our analytical results are overlayed with the 3D micromagnetically-modeled results, using the same material parameters.  We note that the velocities are substantially lower than those predicted previously~\cite{Lemesh2018,Legrand2018} with models that imposed stationary dynamics (dot-dashed lines in Fig.~\ref{fig:7}(a) of Appendix~\ref{sec:extra}) on twisted DWs, emphasizing the qualitative and quantitative impact of precession on $v(j)$. 

\begin{figure}
	\includegraphics[width=1.0\linewidth]{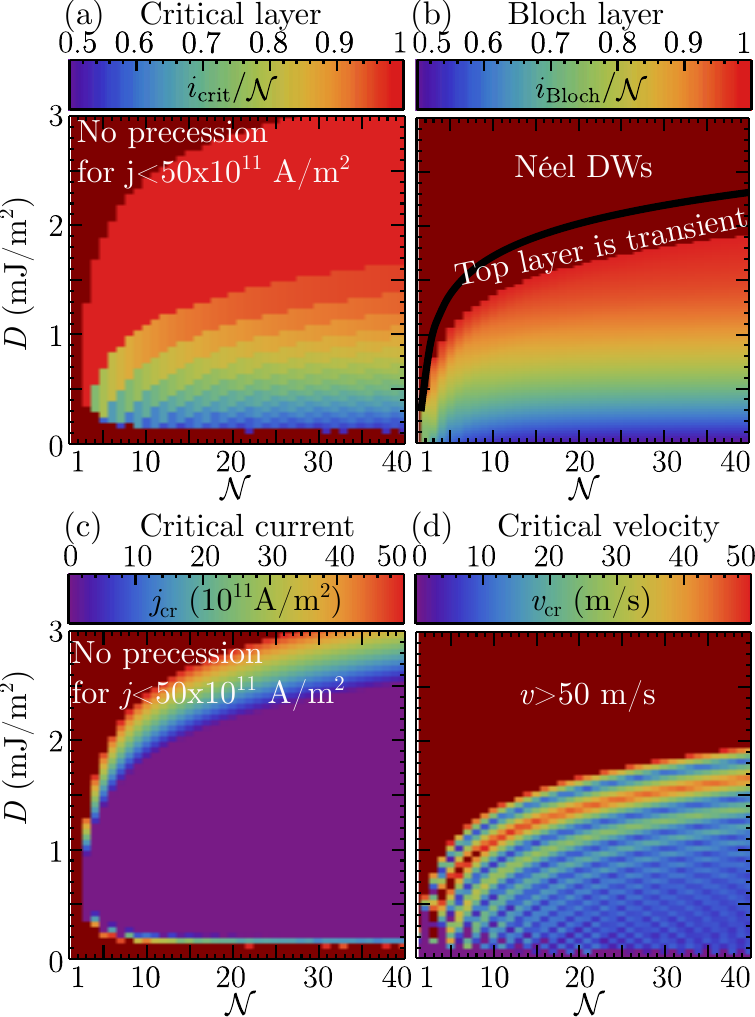}
	\caption{{\bf Critical parameters as a function of DMI constant $D$ and number of multilayer repeats $\mathcal{N}$.} {\bf(a)} Critical layer number $i_{cr}$ for the precession onset and {\bf(b)} the Bloch layer number $i_{\text{Bloch}}$~\cite{Lemesh2018}. {\bf(c)} Critical current $j_{cr}$ for the onset of the first Walker breakdown event and {\bf(d)} the corresponding critical velocity $v_{cr}$. Film parameters are for a film with $f=1/6$, $Q=1.4$, $\alpha=0.3$. Brown color in Figs.~(a),(c) separates regions for which the critical current (if exists) exceeds $j=\SI{50e11}{A/m^2}$, in Fig.~(d) the brown color indicates regions with $v_{cr}$ higher than $\SI{50}{m/s}$, and in Fig.~(b) it indicates regions with no Bloch layers.}\label{fig:5}
\end{figure}

We generally find that Walker breakdown tends to start in the layer whose DW profile is the closest to being Bloch. This is evident from Fig.~\ref{fig:5}(a), which depicts the location of the Walker layer, $i_{cr}$, as a function of the number of multilayer repeats $\mathcal{N}$ and DMI.  Comparing with Fig.~\ref{fig:5}(b), which depicts the location of the Bloch layer~\cite{Lemesh2018} (if it exists), we can conclude that the correlation between $i_{cr}$ and $i_{\text{Bloch}}$ is indeed very high. There are two notable differences: (i) at no DMI, there is no Walker breakdown in the system as it is completely immobile and (ii) at high DMI all the DWs are N\'eel. However, in the latter case, the precession can still occur, although it would always initiate in the topmost layer. This is a consequence of the fact that at high DMI, high currents have the largest impact on the DW angle for a layer that has the smallest $D_{\text{eff}}$, i.e., the one that is near the top of the multilayer stack. Once its DW angle deviates sufficiently from the N\'eel-like configuration, precession ensues. One can thus expect that for low DMI, precession starts close to the middle of the multilayer (approaching $i_{cr}=\mathcal{N}/2$ as $D\rightarrow 0$), while for high DMI, it always begins in the top layer, $i_{cr}=\mathcal{N}-1$ (or $i_{cr}=0$ for negative DMI).

Figure~\ref{fig:5}(c) depicts the critical current as a function of DMI and $\mathcal{N}$. We see that $j_{cr}$ diverges for very small and large values of DMI, with the effect being more dominant for smaller $\mathcal{N}$. For higher $\mathcal{N}$, $j_{cr}(D)$ generally exhibits a wide plateau. The corresponding critical velocity is depicted in Fig.~\ref{fig:5}(d), from which one can find that unless the DMI is very strong, the critical velocity of the DW remains more or less constant. This suggests that $v_{cr}$ is largely independent of the position of the Bloch layer within the multilayer.  This is reasonable, since the Bloch layer is defined by a (near) vanishing of $D_{\text{eff}}$, and the restoring torque in this layer should be the same regardless of its position in a multilayer. This is in general agreement with the established model for Walker breakdown~\cite{Malozemoff1979}, according to which a spin texture exhibits a nonuniform precession once its velocity exceeds a certain critical value.

\begin{figure}
	\includegraphics[width=1.0\linewidth]{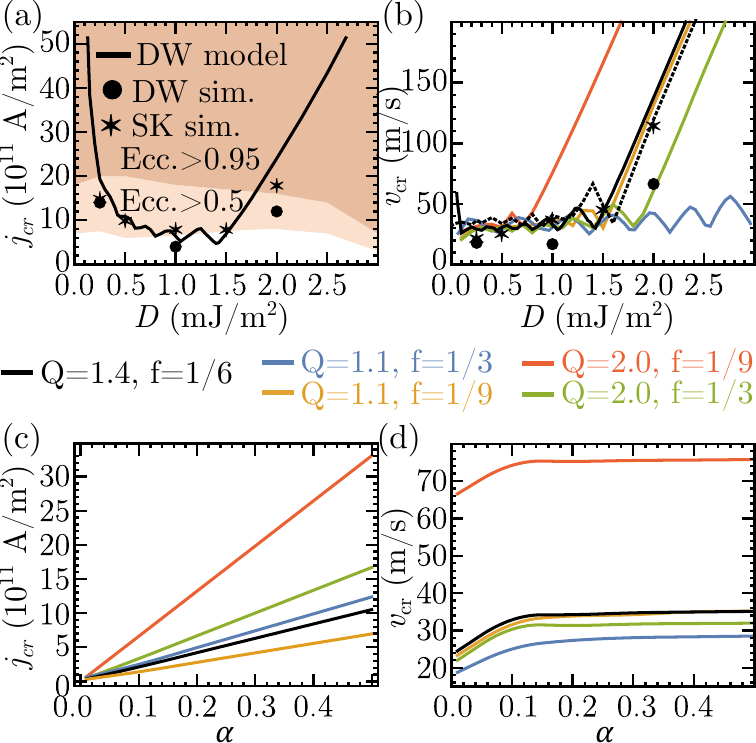}
	\caption{ {\bf(a)} Critical current $j_{\text{cr}}$ and {\bf(b)} critical velocity $v_{\text{cr}}$ as a function of DMI for $\alpha = 0.3$. ({\bf(c)} Critical current $j_{\text{cr}}$ and {\bf(d)} critical velocity $v_{\text{cr}}$ for $D = \SI{1.0}{mJ/m^2}$ as a function of $\alpha$. Results in panels (b)–(d) are shown for several values of the quality factor $Q$ and scaling factor $f$. Continuous lines represent the results obtained from the analytical model, and discrete symbols represent the results of 3D micromagnetic simulations of straight DWs (circles) and skyrmions (stars). Dotted lines in  Fig.~(b) represent $v_{cr}$ given by our simplified model (Eqs.~\eqref{eq:ourwalker},~\eqref{eq:ourfieldwalker}). The color shading in panel (a) indicates the map of skyrmion eccentricity. In all figures, $\mathcal{N}=15$.}\label{fig:6}
\end{figure}

Figure~\ref{fig:6} examines the onset of precession in more detail, focusing on multilayers with $\mathcal{N}=15$ (i.e., focusing on a vertical cut in Figs.~\ref{fig:5}(c) and \ref{fig:5}(d)). The critical velocity and current are illustrated in Figs.~\ref{fig:6}(a) and \ref{fig:6}(b) as a function of $D$. One can see that the analytically-computed trends for both $j_{cr} (D)$ and $v_{cr}(D)$ are in very good agreement with micromagnetic simulations (depicted with points for DWs and stars for skyrmions).  $v_{cr}$ is approximately constant up to some value of DMI, wherein the Bloch layer reaches the top of the film.  At this point, $v_{cr}$ increases approximately linearly with DMI, as a consequence of the disappearance of the Bloch-like layers, in which case, it is more difficult to drive the precession.  This corresponds to a behavior akin to that in single-layer films shown previously~\cite{Thiaville2012}, so $v_{\text{cr}}\propto (D-D_{cr})$. We note that the jagged appearance of the analytical calculation (solid lines in Fig.~\ref{fig:6}(a) and \ref{fig:6}(b) and in the corresponding contour plots of Fig.~\ref{fig:5}) result from the fact that $i$ is a discrete variable so that the Bloch-most layer is in general not located exactly at the node in $D_{\text{eff}}$.

In Fig.~\ref{fig:6}(b), we plot the analytical model solutions for $v_{cr} (D)$ for different quality factors, $Q$, and scaling factors, $f$. We find that the only difference between the resulting $v_{cr}(D)$ curves are in the transition point ($D_{cr}$). Both the offset ($v_{0}$) of the curve $v_{cr}(D)$ (when $D<D_{cr}$) and the slope of the $v_{cr} (D)$ curve (when $D>D_{cr}$) remain approximately independent of $f$ and $Q$ (and $\Delta$). We also find that $v_0$ scales linearly with $M_s$, and with $\mathcal{T}$, but is independent of $\mathcal{N}$ (once there are more than a few layers). 

Finally, the critical current and velocity also depend on the Gilbert damping parameter, $\alpha$. As evidenced from Fig.~\ref{fig:6}(c), $j_{cr}(\alpha)$ increases approximately linearly with $\alpha$ (except for very small $\alpha$). The critical velocity, on the other hand, is constant for $\alpha \gtrsim 0.1$, as it is in conventional Walker breakdown, while for smaller $\alpha$, it varies approximately linearly with current. This dependence of $v_{cr}$ on $\alpha$ originates from the dynamical readjustment of all non-precessing layers $\psi_i$ after every $2\pi$ cycle of precession (resulting in DW oscillations and spreaded DW angles as depicted in Fig.~\ref{fig:8}(c) for layers 0 and 1 in a trilayer heterostructure). When the damping is low, this readjustment occurs over a timescale comparable with the precession period, so all the layers contribute to the $\sum_i \dot{\psi_i}$ term of Eq.~\eqref{eq:qdyn3}. At this point, a further decrease of damping leads to a smaller fraction of time that the non-precessing layers spend in a steady state, which leads to a smaller net DW velocity. In contrast, for high damping, this dynamical readjustment becomes essentially instantaneous, so only the critical precessing layer contributes to the $\sum_i \dot{\psi_i}$ term. In this case, since its precession is not a function of $\alpha$, the critical velocity of the DW is independent of $\alpha$.

\section{Simplified model for the precessional onset threshold} 
Since the observed precessional onset in many ways mimicks conventional Walker breakdown in the presence of DMI, we provide a practical and physically-intuitive model for estimating $v_{cr}$, which is valid for $\alpha\gtrsim 0.1$. Similarly to field-driven DW precessional onset (see Appendix~\ref{sec:walkerfield}), the critical velocity can be found from
\begin{align}
v_{cr}=\Delta\gamma \mu_0 H_\text{\text{stiff}}/2\label{eq:ourwalker},
\end{align}
where $H_\text{\text{stiff}}$ is the strength of the effective ``stiffness'' field that is proportional to the energy difference between states with the critical layer having the Bloch and the N\'eel configurations, defined as
 \begin{align}
H_\text{\text{stiff}}&=\left|\sigma_{tot}^{ 1, \mathcal{N}} (\Delta, \psi_i=\psi_{\text{stat}}, \psi_{cr}=0)\notag \right.\\  & \left.  -\sigma_{tot}^{ 1, \mathcal{N}} (\Delta, \psi_i=\psi_{\text{stat}}, \psi_{cr}=k \pi/2\right|/(\mu_0 M_s f/\mathcal{N})\label{eq:ourfieldwalker},
 \end{align}
Here, $\sigma_{tot}^{ 1, \mathcal{N}}$ is taken from Eq.~\eqref{eq:exactstray}, $\psi_{\text{stat}}$ and $\Delta$ are defined from Eqs. (5, 6) in Ref.~\cite{Lemesh2018}, and $k$ is either $+1$ or $-1$ (chosen for maximum $H_\text{\text{stiff}}$). The resulting $v_{cr}(D)$ curve (dotted lines in Fig.~\ref{fig:6}(b) for the case of Q=1.4, f=1/6) closely follows the numerical and micromagnetic results, even though it was estimated purely from static energy considerations. 

Note that Eq.~\eqref{eq:ourfieldwalker} gives a reasonable value of $v_{cr}$ even at $D=0$ ($v_{cr}=\SI{32.3}{m/s}$). Since in this case, the ``critical'' layer is exactly in the middle of the multilayer stack, surface-volume stray field interactions play no role due to symmetry. Indeed, $D_{sv}(i)$ given by Eq.~\eqref{eq:thr2} at point $ i=(\mathcal{N}-1)/2$ is zero~\cite{Lemesh2018}. Thus, since interlayer interactions vanish for the critical layer, the stiffness field can be approximated from the in-plane shape anisotropy field for a single magnetic layer ($\mathcal{N}=1$, $f=1$), i.e.,
\begin{align}
H_{\text{stiff}}\approx \frac{|(\sigma_{tot}^{ 1, 1}(\psi=0)-\sigma_{tot}^{ 1, 1}(\psi=k\pi/2)|}{\mu_0 M_s}\equiv \frac{2K_{\perp}}{\mu_0 M_s},
\end{align}
where $K_{\perp}=\pi\mu_0M_s^2\Delta^2\mathcal{T}^{-1}G_v(\mathcal{T}/2\pi \Delta)$ is the effective transverse anisotropy~\cite{buttner2015}, with $G_v$ defined in Eq.~\eqref{eq:A3} and in Ref.~\cite{Lemesh2018}. In our case, the magnetic layer is ultrathin ($\mathcal{T}<l_{ex}$), so we can use a thin film approximation for $K_{\perp}$ given by Refs.~\cite{Tarasenko1998,buttner2015},
\begin{align}
K_{\perp}=\frac{\ln(2)\mathcal{T}\mu_0M_s^2}{2\pi\Delta}.
\end{align}
Since Eq.~\eqref{eq:ourwalker} is a general expression for the Walker velocity (see Ref.~\cite{Tarasenko1998} and Appendix~\ref{sec:walker}), we can use it to finally express the critical velocity as
\begin{align}
v_{0}\approx \gamma \mu_0M_s\mathcal{T}\ln(2)/(2\pi),
\end{align}
which equals $v_{0}=\SI{34.2}{m/s}$ for our film parameters, in close agreement with the values we observed in Fig.~\ref{fig:6}(b). This expression is valid for multilayers with $|D|<D_{cr}$. 

Once $|D|>D_{cr}$, we can use the high-DMI limit, assuming that all layers are homochiral N\'eel at equilibrium. In this case, the surface-volume interactions are also absent. Ignoring the interlayer interactions with the upper layer, the upper DW can evolve from N\'eel to the Bloch-like configuration only upon reaching $v_{\text{max}}=\pi \gamma D/(2 M_s)$ as discussed in Appendix~\ref{sec:walkerdmi} and in Ref.~\cite{Thiaville2012}. Thus, when $|D|>D_{cr}$ we can finally obtain
\begin{align}
v_{cr}\approx v_{0}+\frac{\pi \gamma }{2M_s}\left(|D|-D_{cr}\right),\label{eq:vcr}
\end{align}
where $D_{cr}$ (which is positive in our notations) can be roughly approximated by $D_{sv}(0)$ given by Eq.~\eqref{eq:thr2} or more accurately, from the static equations (5) and (6) given in Ref.~\cite{Lemesh2018}, wherein one needs to find D that yields $\sin(\psi_{\mathcal{N}-1})=0$. The latter approach gives $D_{cr}=\SI{1.55}{mJ/m^2}$ for our film parameters, so for the exemplary case of $D=\SI{2.0}{mJ/m^2}$, Eq.\eqref{eq:vcr} results in $v_{cr}=\SI{123}{m/s}$, in agreement with simulations and numerical data plotted in Fig.~\ref{fig:6}(b).

\section{Conclusion}
In conclusion, we show that the phenomenon of Walker breakdown is generally expected to occur in multilayer ferromagnetic films with $\mathcal{N}\geq3$ and finite DMI, both for DWs and for stray field skyrmions. It occurs due to the combined effect of complex surface-volume  stray field interactions, interfacial DMI, and SOT. In this current-induced effect, DWs precess with frequencies in the GHz range. Through simple energetic considerations, we find that the critical velocity for precession for twisted DWs and skyrmions is approximately the same as the Walker velocity for field-driven precession of DWs in a single-layer film with the same properties as each layer in the multilayer.  Although damping-like SOT can drive DWs and skyrmions in single-layer films far beyond the Walker velocity without precession owing to the SOT symmetry, when such layers are incorporated into a multilayer with stray field interactions, precession is generally predicted to occur.  

These results have important implications for potential applications of room-temperature skyrmions in racetrack devices.  Although magnetic multilayers of the type treated here have been widely used to demonstrate stable magnetic skyrmions at room temperature ~\cite{Woo2016,Litzius2016,Legrand2017,Hrabec2017,Woo2017}, the critical velocities for precession for typical material parameters are only of order of tens of m/s. This result implies that in ferromagnetic multilayers, even when the DMI is capable of statically stabilizing skyrmions with a well-defined topological charge, the topological properties are ill-defined during translation, even at relatively modest velocities. Our predictions hence have important technological implications for the use of multilayer-based skyrmions in racetrack devices, since the upper limit for uniform translational velocities is quite low. For this reason, the use of ferromagnetic films for such applications is not technologically viable.  However, since it is the stray field interactions that are ultimately responsible for precessional dynamics identified here, our work points to low-magnetization materials such as ferrimangetic and antiferromagnets~\cite{Buttner2018,Caretta2018} as an alternative path toward realizing practical devices. Finally, while this work provides analytical tools to identify material parameters to allow for optimization of skyrmions for such applications, it may also point to new applications, such as current-driven tunable nano-oscillators~\cite{Carpentieri2015,Garcia-Sanchez2016} based on engineered precessional frequencies in skyrmions.

\begin{acknowledgments}
This work was supported by the U.S. Department of Energy (DOE), Office of Science, Basic Energy Sciences (BES) under Award \#DE-SC0012371 (initial development of domain wall models) and by the DARPA TEE program (examination of instabilities in current-driven dynamics).
\end{acknowledgments}

\onecolumngrid
\appendix

\section{Comparisons between analytical and micromagnetic models}\label{sec:extra}
\begin{figure*}
	\includegraphics[width=1.0\linewidth]{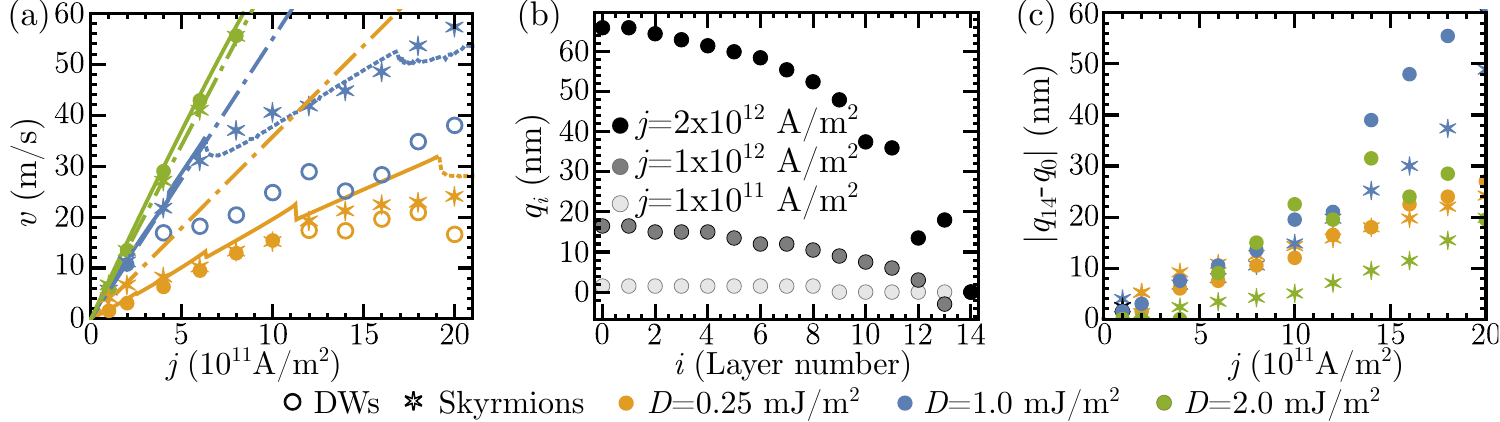}
	\caption{{\bf The effect of current on twisted DWs and skyrmions.} {\bf(a)} Average DW and skyrmion velocity, where continuous (dot-dashed) lines represent the data given by our proposed analytical model (low-current twisted skyrmion theory~\cite{Lemesh2018}), points (stars) represent the data given by micromagnetic simulations for multilayer DWs (skyrmions).  {\bf(b)} The position of DW as a function of layer number and current density for $D=\SI{1.0}{mJ/m^2}$. {\bf(c)} The separation between the top and bottom DWs and skyrmions as a function of current density and DMI. For skyrmions, the decoupling data corresponds to the offsets of their bottom and top centers  (the interlayer radius variation is within $6~\text{nm}$). Magnetic parameters are $\mathcal{N}=15$, $f=1/6$, $Q=1.4$, $\theta_{\text{SH}}=0.1$, $\alpha=0.3$.}\label{fig:7}
\end{figure*}

Here we provide additional micromagnetic simulation results and comparisons to analytical modeling. Figure~\ref{fig:7}(a) shows micromagnetic simulations for isolated DWs (points) and skyrmions (stars). Overlaid are results of the analytical theory provided in the main text (solid and dotted lines), and the results of the static-like twisted skyrmion model presented in Ref.~\cite{Lemesh2018} (dash-dot lines). 

We find that the previous skyrmion theory agrees with simulations only for low currents, which indicates that the profile of a skyrmion remains mostly unchanged upon the current injection (as a consequence of skyrmion rigidity). The exception is $D=\SI{0.25}{mJ/m^2}$, in which case a skyrmion develops a pair of {\it stationary} Bloch lines at $j\approx\SI{4.0e11}{A/m^2}$, which results in a diminished net SOT and hence, lower velocity. Also note that for $D=\SI{0.25}{mJ/m^2}$, the initial slopes of $v(j)$ curves are different for DWs and skyrmions as a consequence of pronounced skyrmion Hall effect at low DMI which leads to a faster velocity than that of a straight DW. 

Above some critical current, the staticlike theory presented previously becomes no longer valid, since in this case, the precessional effects that cannot be captured using a static profile significantly decrease the effective SOT torque and the resulting velocity. At this point, our precessional multilayer DW theory becomes a much better approximation for both simulated DWs and skyrmions. This indicates that the DW precession is a rate-limiting process for both DWs and skyrmions. 

Note that in both models, the agreement is only qualitative. Part of the reason comes from the spatial separation of multilayer DWs through the thickness of the film as a consequence of effective Thiele forces of opposite sign (see Fig.~\ref{fig:1}(b)), which leads to a deviation of the magnetostatic energy from that derived in Ref.~\cite{Lemesh2018} (which relied on the $q_i$=const assumption). This phenomenon of magnetostatic decoupling is demonstrated in Fig.~\ref{fig:7}(b) which depicts the position of DWs as a function of layer number for different current densities. The resulting DW shift is monotonic with layer number and increases with increasing $j$. In Fig.~\ref{fig:7}(c), we depict the maximum shift as a function of current density and DMI. Note that at high currents, this decoupling becomes many times larger than $\Delta$, which leads to an even larger difference between our proposed model and micromagnetic simulations.

\section{Derivation of twisted domain wall dynamics}\label{sec:derivations}

The total volumetric micromagnetic energy density of an isolated DW in a multilayer film can be expressed as
\begin{align}
\mathcal{E}_{tot}^{ 1, \mathcal{N}} & = \mathcal{E}_{\text{exch}}+\mathcal{E}_{\text{DMI}}+\mathcal{E}_{\text{anis}}+\mathcal{E}_{\text{Zeeman}}+\mathcal{E}_{d}^{1, \mathcal{N}} \notag\\&= \frac{1}{\mathcal{N}}\sum_{i=0}^{\mathcal{N}-1} \left\{A  \ \left[\left(\frac{\partial m_{i, x}}{\partial x}\right)^2+\left(\frac{\partial m_{i, y}}{\partial x}\right)^2+\left(\frac{\partial m_{i, z}}{\partial x}\right)^2\right] -D  \left[m_{i, x}\frac{\partial m_{i, z}}{\partial x}-m_{i, z}\frac{\partial m_{i, x}}{\partial x}\right] + K_{u} \left[(m_{i, x})^2+(m_{i, y})^2\right]  \label{eq:exchange0}\notag \right. \\ &\left. -\mu_0M_s (\mathbf{m_i \cdot H})-\frac{1}{2}\mu_0M_s(\mathbf{m_i \cdot H_d})\right\},
\end{align}
where $\mathbf{H}$ is an external magnetic field and $\mathbf{H_{d}}=\mathbf{H_{d}}(x)$ is the demagnetizing field. Here, we follow the index conventions introduced in Ref.~\cite{Lemesh2017,Lemesh2018}, wherein the upper left (right) index indicates the number of DWs (of multilayer repeats) in the system. One way to predict the current- and field-induced evolution of twisted DWs in multilayers is to use the Rayleigh-Lagrange formalism. Similarly to Ref.~\cite{Boulle2011,Boulle2013}, we can introduce the Lagrangian density of the DW (per $x$-$y$  multilayer cross-section area normalized to one layer) as
\begin{align}
\mathcal{L} & =  \int_{-\infty}^{+\infty}\mathcal{E}_{tot}^{ 1, \mathcal{N}}(x)\d x - \frac{1}{\mathcal{N}}\sum_{i=0}^{\mathcal{N}-1} \int_{-\infty}^{+\infty} \d x\left\{\frac{M_sf}{\gamma}\phi_i \sin(\theta_i)\left[\frac{{\d}}{\d t}+u \frac{\d}{\d x}\right]\theta_i -M_sfB_{\text{FL}}\mathbf{m_i\cdot \hat{y}}\right\},\label{eq:lagrangian0}
\end{align}
and the Ralyeigh dissipation functional $\mathcal{F}$ as
\begin{align}
\mathcal{F} = \frac{\alpha M_sf}{2\gamma} \frac{1}{\mathcal{N}}\sum_{i=0}^{\mathcal{N}-1}\int_{-\infty}^{+\infty} \d x\left[\left(\frac{\d}{\d t}-\frac{\beta u}{\alpha} \frac{\d}{\d x}\right)\mathbf{m_i}-\frac{\gamma}{\alpha}B_{\text{DL}}\mathbf{m_i}\times \mathbf{\hat{y}}\right]^2\label{eq:diss0},
\end{align}
where $\gamma=\SI{1.76e11}{sA/kg}$ is the gyromagnetic ratio, $B_{\text{DL}}$ is the damping-like spin orbit torque effective field, $B_{\text{FL}}$ is the field-like spin orbit torque effective field, $u$ is the adiabatic spin-transfer torque parameter (proportional to current density), $\beta$ is the nonadiabatic torque parameter, and $\alpha$ is the damping constant.

Assuming that the ferromagnetic coupling is strong enough to couple domains in all layers, we can use the well-known profile of a DW located at position $q$,
\begin{align}
\theta_i(x, q) & = 2\arctan\{\exp[\mp(x-q)/\Delta]\}, \\\phi_i(t) & = \psi_i(t)-\pi/2,
\end{align}
which corresponds to the following magnetization components:
\begin{align}
m_{i, x} &= \sin(\psi_i) \cosh^{-1}\left(\frac{x-q}{\Delta}\right)\label{eq:magprofilex0},\\
m_{i, y} &= \cos(\psi_i) \cosh^{-1}\left(\frac{x-q}{\Delta}\right)\label{eq:magprofiley0},\\
m_{i, z} &= \pm \tanh\left(\frac{x-q}{\Delta}\right)\label{eq:magprofilez0}.
\end{align}
Here, we also assumed that the DW width is constant in all layers. This assumption, though an approximation~\cite{Lemesh2018,Legrand2017a}, still accurately captures the average width of the DW, $\Delta = \sum_i \Delta_i/\mathcal{N}$~\cite{Lemesh2018}. The total cross-sectional DW energy density $\sigma_{tot}^{ 1, \mathcal{N}} = \int_{-\infty}^{+\infty}\mathcal{E}_{tot}^{1, \mathcal{N}}(x)\d x$ can then be integrated with respect to $x$, which after including the magnetostatic energy of an
infinitely extended ($L_x, L_y \rightarrow \infty$) multilayer film as calculated in Ref.~\cite{Lemesh2018}, can be shown to look as follows:
\begin{align}
\sigma_{tot}^{1, \mathcal{N}}(\Delta, \psi_i) &=\frac{2A}{\Delta}f+2K_{u}\Delta f\pm 2\mu_0H_zM_s f q \notag \\& +\frac{1}{\mathcal{N}}  \sum_{i=0}^{\mathcal{N}-1}\left\{\mp \pi D  f\sin(\psi_i)  -\pi\Delta M_s f\mu_0(H_x\sin(\psi_i)+H_y \cos(\psi_i))\right\} \notag  \\& + \sum_{i=0}^{\mathcal{N}-1}\sum_{j=0}^{\mathcal{N}-1}\left\{F_{s, ij} + \sin(\psi_i)\sin(\psi_{j})  F_{v, ij} \pm \sin(\psi_i)\text{sgn}(i-j)F_{sv, ij}  \right\}\label{eq:E0}.
\end{align}
Here, the generic function $F_{\alpha, ij}$ can be summarized as~\cite{Lemesh2018}
\begin{align}
F_{\alpha, ij} (\mathcal{T},\mathcal{P}, \Delta)=\frac{\pi \mu_{0}M_{s}^2 \Delta^2}{\mathcal{N}\mathcal{P}}\left[ G_{\alpha}\left(\frac{\left|(i-j)\mathcal{P}+\mathcal{T}\right|}{2\pi \Delta}\right)+ G_{\alpha}\left(\frac{\left|(i-j)\mathcal{P}-\mathcal{T}\right|}{2\pi \Delta}\right)  -2G_{\alpha}\left(\frac{\left|(i-j)\mathcal{P}\right|}{2\pi \Delta}\right)\right]\label{eq:A2}
\end{align}
with functions $G_{\alpha}(x)$ defined analytically as follows (for $G_v(x)$):
\begin{align}
G_v(x)&= -2\left\{\Psi^{(-2)}(x+1)-\Psi^{(-2)}\left(x+\frac{1}{2}\right)- x \ \ln(\Gamma(x+1)) 
+x \ln\left[\Gamma\left(x+\frac{1}{2}\right)\right] \right. \notag\\ &\left.- \Psi^{(-2)}(1)+\Psi^{(-2)}\left(\frac{1}{2}\right) 
\right\},\label{eq:A3}\\
G_s(x) &=-\left\{\Psi ^{(-2)}(2 x)+x^2 (2  \log (x)+\log (4)-1)-x(1+2  \ln[\Gamma (2 x)])\right\},\label{eq:A4}\\
G_{sv}(x) & = 2 \ln\left[\Gamma\left(x+\frac{1}{2}\right)\right]\label{eq:A5},
\end{align}
where $\Gamma$ is the gamma function and $\Psi^{(-2)}(z)=\int_0^z \d t \ln\Gamma(t)$ is the second anti-derivative of the digamma function $\Psi$. 

The Lagrangian, Eq.~\eqref{eq:lagrangian0}, and the dissipation function, Eq.~\eqref{eq:diss0}, can also be integrated, resulting in
\begin{align}
\mathcal{L} & = \sigma_{tot}^{ 1, \mathcal{N}}(\Delta, \psi_i) + \frac{1}{\mathcal{N}}\sum_{i=0}^{\mathcal{N}-1} \left\{\pm \frac{2M_sf}{\gamma}\left(\psi_i-\frac{\pi}{2}\right)(\dot{q}-u)+\pi \Delta M_sfB_{\text{FL}}\cos(\psi_i)\right\},\label{eq:L0}
\end{align}
\begin{align}
\mathcal{F} = \frac{\alpha M_sf}{\gamma}\left[\frac{1}{\Delta}\left(\frac{\beta u}{\alpha}-\dot{q}\right)^2+\frac{\pi^2\dot{\Delta}^2}{12\Delta} \right]+\frac{\alpha M_sf}{\gamma} \frac{1}{\mathcal{N}}\sum_{i=0}^{\mathcal{N}-1}\left\{\dot{\psi_i}^2 \Delta \mp \frac{\pi \gamma B_{\text{DL}} }{\alpha}\left(\frac{\beta u}{\alpha}-\dot{q}\right)\sin(\psi_i)-\frac{\Delta \gamma^2 B_{\text{DL}}^2}{\alpha^2} \cos^2(\psi_i)\right\}.\label{eq:F0}
\end{align}

The equations describing the evolution of the DW profile can then be obtained from the Lagrange-Rayleigh equations
\begin{align}
\frac{\partial\mathcal{L}}{\partial q}-\frac{\partial}{\partial t}\frac{\partial\mathcal{L}}{\partial \dot{q}}-\frac{\partial}{\partial x}\frac{\partial\mathcal{L}}{\partial q'}+\frac{\partial\mathcal{F}}{\partial \dot{q}}&=0, \label{eq:q0}\\
\frac{\partial\mathcal{L}}{\partial \psi_i}-\frac{\partial}{\partial t}\frac{\partial\mathcal{L}}{\partial\dot{\psi_i}}-\frac{\partial}{\partial x}\frac{\partial\mathcal{L}}{\partial \psi_i'}+\frac{\partial\mathcal{F}}{\partial \dot{\psi_i}}&=0\label{eq:psid0},\\ \frac{\partial\mathcal{L}}{\partial \Delta}-\frac{\partial}{\partial t}\frac{\partial\mathcal{L}}{\partial \dot{\Delta}}-\frac{\partial}{\partial x}\frac{\partial\mathcal{L}}{\partial \Delta'}+\frac{\partial\mathcal{F}}{\partial \dot{\Delta}}&=0. \label{eq:deltad0}
\end{align}
After substituting $\mathcal{L}$, $\mathcal{F}$ from Eqs.~$\eqref{eq:E0}$,~\eqref{eq:L0}, and \eqref{eq:F0} into Eqs.~\eqref{eq:q0}-\eqref{eq:deltad0}, we can finally obtain
\begin{align}
\pm \frac{\dot{q}\alpha}{\Delta} -  \frac{1}{\mathcal{N}}  \sum_{j=0}^{\mathcal{N}-1} \dot{\psi_j}&=  -\gamma \mu_0 H_z \pm \frac{\beta u}{\Delta} - \frac{\pi \gamma B_{\text{DL}} }{2\mathcal{N}}  \sum_{j=0}^{\mathcal{N}-1}\sin(\psi_j), \label{eq:qdyn0} \\
\mp \frac{\dot{q}}{\Delta}- \alpha\dot{\psi_i} &=\mp  \frac{u}{ \Delta}+\frac{\pi\gamma}{2} \cos(\psi_i)\left(\mp \frac{D}{M_s \Delta}-\mu_0H_x\right)+\frac{\pi \gamma}{2} \sin(\psi_i)(\mu_0H_y-B_{\text{FL}})\notag\\& + \frac{\gamma \mathcal{N} \cos(\psi_i)}{2  M_s f \Delta} \sum_{j=0}^{\mathcal{N}-1}\left[(1+\delta_{ij})\sin(\psi_{j})  F_{v, ij}(\Delta) \pm  \text{sgn}(i-j)F_{sv, ij}(\Delta)\right],\label{eq:psidyn0}\\
\dot{\Delta}& = \frac{6\Delta\gamma}{\pi^2\alpha M_s f}\left\{\frac{2Af}{\Delta^2}-2K_uf+\pi M_sf \frac{1}{\mathcal{N}}  \sum_{i=0}^{\mathcal{N}-1}\left[\mu_0H_x\sin(\psi_i)+ (\mu_0 H_y-B_{\text{FL}})\cos(\psi_i)\right]\right. \notag\\ &\left.  -  \sum_{j=0}^{\mathcal{N}-1}\sum_{i=0}^{\mathcal{N}-1} \left[\sin(\psi_i)\sin(\psi_{j}) \frac{\partial F_{v, ij} (\Delta) }{\partial \Delta }  \pm    \sin(\psi_i)\text{sgn}(i-j)\frac{\partial F_{sv, ij}(\Delta)}{\partial \Delta } +  \frac{\partial F_{s, ij}(\Delta) }{\partial \Delta }\right]\right\}.\label{eq:deltadyn0}
\end{align}
By solving Eqs.~\eqref{eq:qdyn0}, \eqref{eq:psidyn0}, and \eqref{eq:deltadyn0} simultaneously one can extract the time-dependent  ($q$, $\psi$, $\Delta$), i.e., reveal the evolution of the twisted DWs in magnetic multilayers. Below, we introduce a few further simplifications to these equations.

For a freely propagating DW, we can combine Eq.~\eqref{eq:qdyn0} and Eq.~\eqref{eq:psidyn0} to eliminate the $q$-dependence, which leads to $\mathcal{N}$ additional equations for each layer:
\begin{align}
\alpha^2\dot{\psi_i} +  \frac{1}{\mathcal{N}}  \sum_{j=0}^{\mathcal{N}-1} \dot{\psi_j}&=  \gamma \mu_0 H_z \pm \frac{u}{ \Delta} (\alpha -\beta) +\frac{\pi \gamma}{2} \left(-\alpha \sin(\psi_i)(\mu_0H_y-B_{\text{FL}}) +\frac{B_{\text{DL}} }{\mathcal{N}}  \sum_{j=0}^{\mathcal{N}-1}\sin(\psi_j) \right)  \notag\\& + \frac{\alpha \gamma \cos(\psi_i)}{2M_s\Delta} \left(\pm \pi D+\mu_0H_x M_s\Delta - \frac{\mathcal{N}}{  f } \sum_{j=0}^{\mathcal{N}-1}\left[(1+\delta_{ij})\sin(\psi_{j})  F_{v, ij}(\Delta) \pm  \text{sgn}(i-j)F_{sv, ij}(\Delta)\right]\right)\label{eq:psidyn20}
\end{align}
Let us consider the case in which the width of the DW remains constant, $\dot{\Delta}=0$. Then from Eq.~\eqref{eq:deltadyn0} we can derive the following equation for the equilibrium $\Delta$,
\begin{align}
2K_uf-\frac{2Af}{\Delta^2}&+\pi M_sf \frac{1}{\mathcal{N}}  \sum_{i=0}^{\mathcal{N}-1}[ (B_{\text{FL}}-\mu_0H_{y})\cos(\psi_i) -\mu_0H_x\sin(\psi_i)]\notag \\ & +  \sum_{j=0}^{\mathcal{N}-1}\sum_{i=0}^{\mathcal{N}-1} \left[\sin(\psi_i)\sin(\psi_{j}) \frac{\partial F_{v, ij} (\Delta) }{\partial \Delta }  \pm   \sin(\psi_i)\text{sgn}(i-j)\frac{\partial F_{sv, ij}(\Delta)}{\partial \Delta } +  \frac{\partial F_{s, ij}(\Delta) }{\partial \Delta }\right]=0\label{eq:deltadyn20}
\end{align}
Note that with the exception of the $B_{\text{FL}}$, $H_{x}$,  and $H_{y}$ terms, this equation is identical to the static equation for the equilibrium $\Delta$ (see Eq.~(6) in Ref.~\cite{Lemesh2018}). Finally, the velocity of the DW can be found from Eq.~\eqref{eq:qdyn0} as
\begin{align}
\dot{q}&= \frac{\beta u}{\alpha} \mp\frac{ \Delta}{\alpha}\left(\gamma \mu_0H_z + \sum_{j=0}^{\mathcal{N}-1}\left[\frac{\pi \gamma B_{\text{DL}} }{2}  \sin(\psi_j)- \dot{\psi_j}\right]\right)\label{eq:qdyn20} 
\end{align}

\section{The presence of Walker breakdown}\label{sec:walker}
Let us now derive the criterion that can help us predict the presence or absence of DW precession. For this,  consider the steady state equation, which can be found by setting $\dot{\psi_i}=0$ in Eqs.~\eqref{eq:psidyn20}-\eqref{eq:qdyn20}. For simplicity, consider that only the damping-like SOT and out-of-plane magnetic field are present, in which case we obtain
\begin{align}
& \left(\sum_{j=0}^{\mathcal{N}-1}\left[(1+\delta_{ij})\sin(\psi_{j})  F_{v, ij}  \pm  \text{sgn}(i-j)F_{sv, ij}\right]\mp \frac{\pi  f}{\mathcal{N}} D\right) \cos(\psi_i) -\frac{2 M_s f \Delta}{\alpha \mathcal{N} }\left[\frac{\pi  B_{\text{DL}}}{2 \mathcal{N}} \sum_{j=0}^{\mathcal{N}-1} \sin(\psi_j)+\mu_0H_z\right]=0 \label{eq:psidyn4},\\
\dot{q}&= \mp\frac{ \Delta}{\alpha}\left(\gamma \mu_0 H_z +\frac{\pi \gamma B_{\text{DL}} }{2 \mathcal{N}} \sum_{j=0}^{\mathcal{N}-1}\left[  \sin(\psi_j)\right]\right)\label{eq:qdyn4} 
\end{align}
The Walker breakdown is present whenever one can find such fields or currents, for which Eq.~\eqref{eq:psidyn4} yields no real $\psi_i$ solutions~\cite{Linder2013,Risinggard2017}.

\subsection{Conventional Walker breakdown}\label{sec:walkerfield}
First, we review conventional Walker breakdown~\cite{Malozemoff1979} occurring in single layer films ($\mathcal{N}=1$, $f=1$) in the presence of an easy-axis magnetic-field $B_z$ and no DMI. Eqs.~\eqref{eq:psidyn4},~\eqref{eq:qdyn4} then reduce to
\begin{align}
& 2\sin(\psi_{0})\cos(\psi_0)  F_{v, 00} -2 M_s  \Delta \mu_0 H_z/\alpha =0 \label{eq:psidyn5}\\
\dot{q}&=  \mp \Delta\gamma \mu_0 H_z/\alpha \label{eq:qdyn5},
\end{align}
which clearly yields an analytical solution for the DW angle, $\sin(2\psi_0)=2M_s\Delta \mu_0 H_z/(\alpha F_{v, 00})$. For this reason, the DW angle ranges from $\psi_0=\pi n$ (Bloch state) at zero field to $\pi/4+\pi n$ at some critical field $H_{cr}=\alpha F_{v, 00}/(2\mu_0 M_s\Delta)$, all of which correspond to steady state solutions. However, for $H_z>H_{cr}$ precessional motion occurs, since in this case, there is no real steady-state solution of Eq.~\eqref{eq:psidyn5}. For this reason, $H_{cr}$ is also known as the Walker critical field, which is more commonly expressed as $H_{cr}=\alpha H_K/2$, where $H_K=2 K_{\perp}/\mu_0M_s$ is the in-plane anisotropy field~\cite{Malozemoff1979}. Here, $K_{\perp}$ is the effective transverse anisotropy, which using Eqs.~\eqref{eq:psidyn5},~\eqref{eq:A2},~\eqref{eq:A3} can be expressed as $K_{\perp}=\pi\mu_0M_s^2\Delta^2\mathcal{T}^{-1}G_v(\mathcal{T}/2\pi \Delta)$~\cite{buttner2015}. Note that in order to start precessing, the DW needs to be driven to the {\it Walker breakdown velocity}, which from Eq.~\eqref{eq:qdyn5} is 
\begin{align}
v_{cr}=\Delta\gamma \mu_0 H_K/2.\label{eq:conventionalwalker}
\end{align}

\subsection{Absence of precession for single layer films driven by damping-like SOT}\label{sec:walkerdmi}
Consider a film with DMI and damping-like SOT. Starting from a single magnetic layer ($\mathcal{N}=1$, $f=1$), Eqs.~\eqref{eq:psidyn4},~\eqref{eq:qdyn4} can be expressed as
\begin{align}
& \sin(2\psi_{0})  F_{v, 00}  \mp \pi D\cos(\psi_0) = \sin(\psi_0)\frac{\pi M_s \Delta B_{\text{DL}}}{\alpha} \label{eq:psidyn6},\\
&\dot{q}=  \mp \pi\gamma\Delta B_{\text{DL}}\sin(\psi_0)/(2\alpha) \label{eq:qdyn6},
\end{align}
where from Eqs.~\eqref{eq:A2},~\eqref{eq:A3}, $F_{v, 00}=2\pi\mu_0M_s^2\Delta^2 \mathcal{N}^{-1}\mathcal{T}^{-1} G_v(\mathcal{T}/2\pi \Delta)$.

\begin{figure*}
	\includegraphics[width=1.0\linewidth]{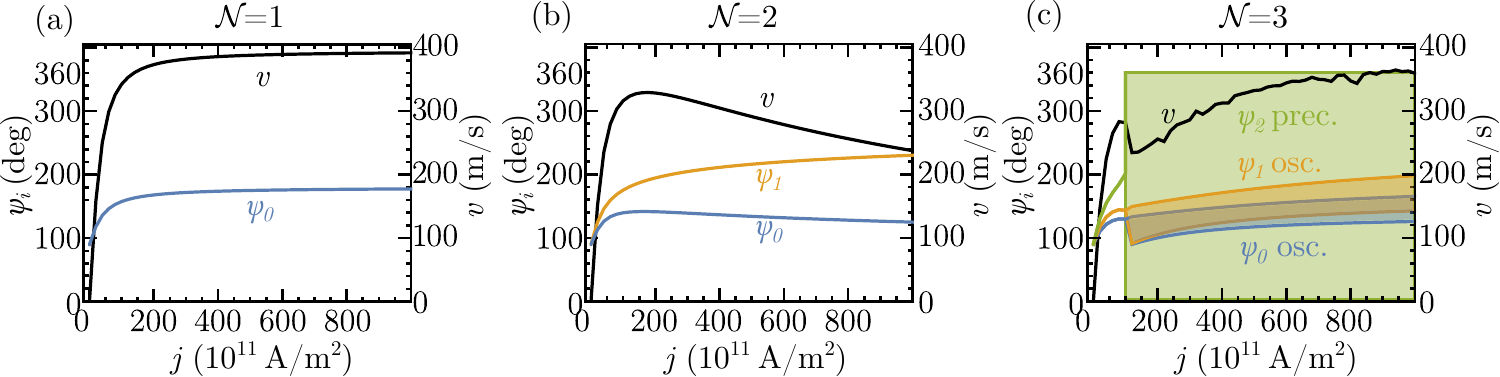}
	\caption{{\bf Numerical solution of dynamic Eqs.~\eqref{eq:psidyn20}-\eqref{eq:qdyn20} for the SOT-driven motion}. Films have {\bf(a)}  $\mathcal{N}=1$, {\bf(b)}  $\mathcal{N}=2$, {\bf(c)} $\mathcal{N}=3$ multilayer repeats. All the cases depict a steady state dynamics that can also be identically  described by Eqs.~\eqref{eq:psidyn4},~\eqref{eq:qdyn4}, with the exception of Figure {\bf(c)}, which also depicts the precession and oscillation of DWs for $j>j_{cr}=\SI{1.03e13}{A/m^2}$. Material and external parameters are  $D=\SI{2.0}{mJ/m^2}$ (at $j=0$, all DWs are N\'eel with $\psi_i=+\pi/2$), $\theta_{SH}=+0.1$, $\alpha=0.3$, $f=1/6$, $Q=1.4$, $H_z=0$.}\label{fig:8}
\end{figure*}

First, from Eq.~\eqref{eq:qdyn6} we can see that even if the film had a pure Bloch wall state ($\psi_0=\pi n$), the DW would be completely immobile, because of the $\sin(\psi_0)$-dependence of velocity. As for non-Bloch DWs, Eq.~\eqref{eq:psidyn6} can be expressed as 
\begin{align}
\sin(\psi_0)= \frac{\pm\pi D\cos(\psi_0)}{\pi M_s \Delta B_{\text{DL}}/\alpha -2\cos(\psi_0)F_{v, 00}}\label{eq:singlesteady}
\end{align}
One can easily find with numerics that for any $B_{\text{DL}}$ there always exists a {\it real-valued} solution for $\psi_0$. This implies a universal absence of precession~\cite{Linder2013,Risinggard2017}, so the DW is always at a steady state, with the profile ranging between the N\'eel state, $\psi_0=\pm \pi/2 \text{sgn}(D)$  (or transient $\psi_0=\pm\arcsin(\pi D/2F_{v, 00})$ for intermediate DMI and $\mathcal{T}$~\cite{Lemesh2017}) at no current, and approaching the Bloch-like state $\psi_0\rightarrow\pm\pi /2[1+\text{sgn}(B_{\text{DL}})]\text{sgn}(D)$ at very high current. These limiting cases are clearly evident from Fig.~\ref{fig:8}(a), which depicts the solution of Eq.~\eqref{eq:singlesteady} for $D=\SI{2.0}{mJ/m^2}$. 
In the case of a high current, one can use a Taylor expansion for $\psi_0$ in Eq.\eqref{eq:singlesteady}, which from Eq.\eqref{eq:qdyn6} then results in the maximum velocity
\begin{align}
v_{\text{max}}=\mp \frac{\pi \gamma D}{2 M_s}\label{eq:singlesdmi},
\end{align}
which is independent of current, as expected for films with no STT, but finite damping-like SOT and DMI~\cite{Risinggard2017}. Note that single layer ferromagnets still exhibit a finite Walker velocity limit when driven with the SOT, although in this case it serves as an upper limit that can be reached only asymptotically.

\subsection{Absence of precession for bilayers}
We now consider a magnetostatically coupled bilayer film ($\mathcal{N}=2$), which corresponds to an asymmetric H/M/S/H/M/S-type heterostructure (rather than to a symmetric H/M/S/M/H bilayer~\cite{Hrabec2017}, for which the Thiele forces add up constructively).  Numerical solution of Eq.~\eqref{eq:psidyn20} indicates that the DW is always in a steady state, which implies a universal absence of Walker breakdown for such ferromagnetic bilayers (as was also found for the exchange-coupled bilayers of compensated synthetic antiferromagnets~\cite{Risinggard2017}). Note that unlike in the single-layer case, the trends for $\psi_1$, $\psi_2$ and $v$ are non-monotonic with current (as visualized for $D=\SI{2.0}{mJ/m^2}$ in Fig.~\ref{fig:8}(b)). At small currents, at least one of the layers has a N\'eel DW orientation. As the current increases, the structure, first, reaches some maximum velocity (with neither of the DWs being N\'eel), while at very high currents, the velocity drops to zero, with the structure stabilizing to a terminal configuration (with the N\'eel walls of opposite chiralities as we see next). 

We can verify the absence of precession by analyzing the steady-state solutions given by Eqs.~\eqref{eq:psidyn4},~\eqref{eq:qdyn4},
\begin{align}
\pi f\Delta M_s B_{\text{DL}}[\sin(\psi_0)+\sin(\psi_1)]/(4\alpha)&=\cos(\psi_0)[2\sin(\psi_0)F_{v, 00}+\sin(\psi_1)F_{v, 01}\mp (F_{sv, 01}+\pi D f/2)]\label{eq:qdyn70}\\
&=\cos(\psi_1)[\sin(\psi_0)F_{v, 01}+2\sin(\psi_1)F_{v, 00} \pm (F_{sv, 01}-\pi D f/2)],\label{eq:qdyn71}\\
\dot{q}=  \mp \pi\gamma\Delta &B_{\text{DL}}[\sin(\psi_0)+\sin(\psi_1)]/(4\alpha),\label{eq:psidyn7}
\end{align}
where we have also used the fact that both matrices $[F_{v, ij}]$ and $[F_{sv, ij}]$ are centrosymmetric, with the values of their elements depending only on $|i-j|$~\cite{Lemesh2018}. Using numerics, one can verify that solving Eqs.~\eqref{eq:qdyn70},~\eqref{eq:qdyn71} always leads to real valued $\psi_0$ and $\psi_1$, indicating a steady state for all $j$. We can examine the limiting cases analytically. The scenario of $j\rightarrow 0$ trivially reduces to the static solutions covered in Ref.~\cite{Lemesh2018}. As for $j\rightarrow\infty$ (i.e., for $B_{\text{DL}}\rightarrow\infty $), the only possibility for the system of Eqs.~\eqref{eq:qdyn70},~\eqref{eq:qdyn71} to yield real-valued solutions is to have 
\begin{align}
\sin(\psi_0)+\sin(\psi_1)= C_2/B_{\text{DL}}+\mathcal{O}(1/B_{\text{DL}}^2),\label{eq:eq200}
\end{align}
where $C_2$ is some constant. Hence, from Eqs.~\eqref{eq:qdyn70},~\eqref{eq:qdyn71}, we obtain
\begin{align}
\frac{\pi f\Delta M_s C_2}{4\alpha}
& \approx \cos(\psi_0)[\sin(\psi_0)(2F_{v, 00}-F_{v, 01})\mp\pi (D f/2 + F_{sv, 01})],\label{eq:eq20}\\
&\approx \cos(\psi_1)[\sin(\psi_1)(2F_{v, 00}-F_{v, 01}) \mp\pi (D f/2 - F_{sv, 01})].\label{eq:eq21}
\end{align}
One can always find some real $C_2$ that results in the simultaneous solution of both of Eq.~\eqref{eq:eq20} and Eq.~\eqref{eq:eq21}, such that $\sin(\psi_0)+\sin(\psi_1)\approx0$ (from Eq.~\eqref{eq:eq200}). Thus, steady state solutions exist even at very large currents. A simple assymptotic analysis of Eqs.~\eqref{eq:eq200},~\eqref{eq:eq20},~\eqref{eq:eq21} at $B_{\text{DL}}\rightarrow\infty$ yields $\psi_0  \rightarrow \pm \pi/2 $ and $ \psi_1\rightarrow \mp \pi/2$, and $C_2\rightarrow 0$. From Eq.~\eqref{eq:psidyn7}, this corresponds to the terminal velocity of $v_{\text{term}}=  \mp \pi\gamma\Delta C_2/(4\alpha)\rightarrow 0$. 

Thus, the bilayer system has no means to reach the critical Walker velocity at very high currents. It reaches the maximum velocity at some intermediate values of current, though in this case, numerical solution of Eqs.~\eqref{eq:qdyn70},~\eqref{eq:qdyn71} always yields some steady state solutions, indicating a universal absence of precession (for any $j$). These results remain valid even for films with $D_1\neq D_2$ and/or for finite $H_x$, $H_y$ fields. In the light of the $N\geq3$ case that we consider below, we attribute such absence of precession to the high symmetry of bilayers, which leads to an insufficient complexity of stray field interactions.

\subsection{Walker breakdown of trilayers and $\mathcal{N}>3$ multilayers}
Let us now consider $\mathcal{N}\geq3$ multilayers. By resolving Eq.~\eqref{eq:psidyn20} numerically, we can find that once the current exceeds a certain threshold, DW always starts to precess. Let us focus on a magnetic trilayer and demonstrate this explicitly by analyzing the steady state Eqs.~\eqref{eq:psidyn4},~\eqref{eq:qdyn4}, for which we can follow a similar logic as for bilayers and express the steady state as
\begin{align}
\pi f\Delta M_s B_{\text{DL}}&[\sin(\psi_0)+\sin(\psi_1)+\sin(\psi_2)]/(9\alpha)=\notag\\=\cos(\psi_0)&[2\sin(\psi_0)F_{v, 00}+\sin(\psi_1)F_{v, 01}+\sin(\psi_2)F_{v, 02}\mp (F_{sv, 01}+F_{sv, 02}+\pi D f/3)],\label{eq:qdyn80}\\
=\cos(\psi_1)&[\sin(\psi_0)F_{v, 01}+2\sin(\psi_1)F_{v, 00}+\sin(\psi_2)F_{v, 01} \mp \pi D f/3],\label{eq:qdyn81}\\
=\cos(\psi_2)&[\sin(\psi_0)F_{v, 02}+\sin(\psi_1)F_{v, 01}+2\sin(\psi_2)F_{v, 00} \pm (F_{sv, 01}+F_{sv, 02}-\pi D f/3)],\label{eq:qdyn82}\\
\dot{q} &=  \mp \pi\gamma\Delta B_{\text{DL}}[\sin(\psi_0)+\sin(\psi_1)+\sin(\psi_2)]/(6\alpha).\label{eq:psidyn8}
\end{align}
At small currents, this system can be resolved, yielding the real-valued steady state solutions (the case of $j\rightarrow 0$ is trivial as it reduces to the static solutions covered in Ref.~\cite{Lemesh2018}). However, for large currents, these equations are irresolvable, as the DW precession takes over. In this case, the corresponding dynamic solutions can be obtained only by resolving Eqs.~\eqref{eq:psidyn20}-\eqref{eq:qdyn20}, as visualized for $D=\SI{2.0}{mJ/m^2}$ in Fig.~\ref{fig:8}(c).

\begin{figure}
	\includegraphics[width=1.0\linewidth]{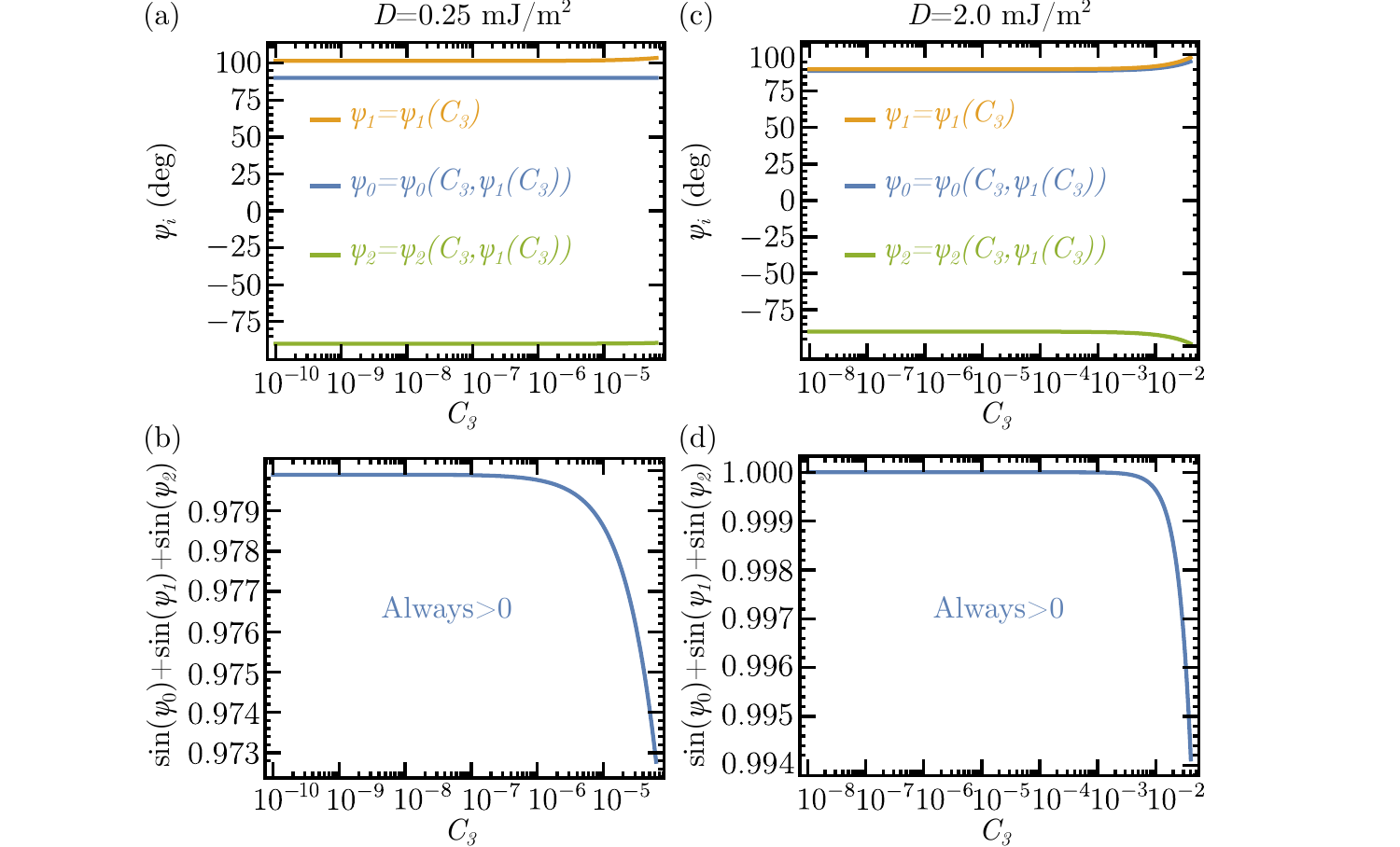}
	\caption{{\bf Numerical solution of Eqs.~\eqref{eq:three1}-\eqref{eq:three3}.} {\bf(a), \bf(b)} for $D=\SI{0.25}{mJ/m^2}$ and {\bf(c), \bf(d)} for $D=\SI{2.0}{mJ/m^2}$. For all values of $C_3$, the expression $\sin(\psi_0)+\sin(\psi_1)+\sin(\psi_2)\approx 0$ (Eq.~\eqref{eq:three0}) cannot be satisfied. Note that the domains of functions $\psi_i(C_3)$ are constrained from above by $C_3$, so the range is plotted in accordance with it. $\mathcal{N}=3$, $Q=1.4$, $f=1/6$, $\alpha=0.3$.}\label{fig:9}
\end{figure}
Similarly to the $\mathcal{N}=2$ case, the only possibility for each of  Eqs.~\eqref{eq:qdyn80},~\eqref{eq:qdyn81},~\eqref{eq:qdyn82} to yield real-valued solutions at $j\rightarrow\infty$ (i.e., at $B_{\text{DL}}\rightarrow\infty $) is to have
\begin{align}
\sin(\psi_0)+\sin(\psi_1)+\sin(\psi_2)= C_3/B_{\text{DL}}+\mathcal{O}(1/B_{\text{DL}}^2),\label{eq:three0}
\end{align}
so Eqs.~\eqref{eq:qdyn80},~\eqref{eq:qdyn81},~\eqref{eq:qdyn82}  can be approximated as
\begin{align}
\pi f\Delta M_s C_3/(9\alpha)&\approx
\cos(\psi_1)[\sin(\psi_1)(2F_{v, 00}-F_{v, 01}) \mp \pi D f/3], \label{eq:three1}\\
\approx&\cos(\psi_0)[\sin(\psi_0)(2F_{v, 00}-F_{v, 02})+\sin(\psi_1)(F_{v, 01}-F_{v, 02})\mp (\pi D f/3+F_{sv, 01}+F_{sv, 02})],\label{eq:three2}\\
\approx&\cos(\psi_2)[\sin(\psi_1)(F_{v, 01}-F_{v, 02})+\sin(\psi_2)(2F_{v, 00}-F_{v, 02}) \mp (\pi D f/3 -F_{sv, 01}-F_{sv, 02})],\label{eq:three3}
\end{align}
We verify with numerics that Eqs.~\eqref{eq:three0}-\eqref{eq:three3} possess no mutual solutions (as shown in Fig.~\ref{fig:9}), regardless of the value of $C_3$. Hence, at infinite current, the system must have non-steady (precessing) solutions. This holds true also for multilayers with $\mathcal{N}>3$, since the complexity of the corresponding system of equations can only increase with $\mathcal{N}$.  In contrast, removing the $F_{sv}$ terms removes the dominant twist, which leads to the trivial layer-independent solutions $\psi_i\rightarrow\pm\pi /2[1+\text{sgn}(B_{\text{DL}})]\text{sgn}(D)$ at $j\rightarrow\infty$, as in the single layer case. For this reason, the observed Walker breakdown originates from the surface-volume stray field interactions. These interactions tend to reduce the restoring torque in some layers, hence lowering the Walker threshold velocity for those layers. Even though the SOT acting on these layers is by itself insufficient to drive them to the critical velocity, the Thiele forces acting on the composite structure are able to do so.  Therefore, the Bloch-like wall can be driven to the Walker velocity and start to precess.

Hence, SOT-driven ferromagnetic multilayers with $\mathcal{N}=1, 2$  always exhibit a universal absence of Walker breakdown, but a finite critical current for precession exists generally for $\mathcal{N}>2$ as long as the DMI is finite.

\subsection{Walker breakdown in synthetic ferrimagnet bilayers}
The absence of Walker breakdown has also been proved for {\it exchange coupled} DWs in PMA synthetic antiferromagnet (SAF) stacks with stacking structure H0/M/S/M/H1,  as discussed in Ref.~\cite{Risinggard2017}. However, this is no longer true if the magnetic layers that comprise SAF possess different saturation magnetizations, $M_{s0}\neq M_{s1}$ (i.e., they constitute a H0/M0/S/M1/H1-type synthetic $\it ferri$magnet heterostructure). In this case, as was described in a recent study~\cite{Yang2019}, the interplay of negative interlayer exchange coupling and sufficient negative field $H_x$ can also lead to the phenomenon of domain wall precession. 

Note that in the light of this paper and our earlier work~\cite{Lemesh2018}, the treatment of magnetostatics by Ref.~\cite{Yang2019}, and hence, of the resulting equation of DW motion can be improved even further. According to Ref.~\cite{Yang2019} (see supplemental Eq.~S14), the integrated dipolar energy (here, per unit length along the $y-$direction) of the DW in a SAF structure can be expressed as
\begin{align}
\lambda_d = \mu_0 M_L H^k_L \Delta_L \cos^2(\psi_L) + \mu_0 M_U H^k_U \Delta_U \cos^2(\psi_U)+E^{dip}_{L}\Delta_L+E^{dip}_{U}\Delta_U
\end{align}
where $\Delta_i\approx\sqrt{A_i/(K_i+E_{i}^{dip})}$~\cite{Yang2019} and $M_L$ and $M_U$ are magnetizations multiplied by the corresponding film thicknesses. Using our notations and DW angle definition, this corresponds to 
\begin{align}
\lambda_{d}^{1, 2 ({\text{SAF}})}  = \mu_0 M_{s, 0} \mathcal{T}_0 H_{k, 0} \Delta_0 \sin^2(\psi_0) + \mu_0 M_{s, 1} \mathcal{T}_1 H_{k, 1} \Delta_1 \sin^2(\psi_1)+E^{dip}_{0}\Delta_0+E^{dip}_{1}\Delta_1 \label{app:yang}
\end{align}
As we demonstrate below, while this treatment is a reasonable approximation for the {\it intra}layer dipolar interactions, it completely ignores the mutual {\it inter}layer interactions.

From our earlier work~\cite{Lemesh2018}, the total magnetostatic energy (per single DW area) of bilayer DWs with different $M_{s, i}$ and $\Delta_i$ (but the same $\mathcal{T}_1=\mathcal{T}_2\equiv \mathcal{T}$) has the form
\begin{align}
\sigma_{d}^{1, 2 ({\text{SAF}})}  = \sigma_{d, s}^{1, 2 ({\text{SAF}})} + \sigma_{d, v}^{1, 2 ({\text{SAF}})} + \sigma_{d, sv}^{1, 2 ({\text{SAF}})}  
\end{align}
In the SAF structure, these components can be expressed by assuming that the profile of the upper DW has $\theta_1(x) = \arctan\{\exp[\mp (x-q_i)/\Delta_1]\}$ and $\psi_1$, while the lower DW has $\theta_0(x) = \arctan\{\exp[\mp\text{sign}(J^{ex})(x-q_i)/\Delta_0]\}$ and $\psi_0$~\cite{Yang2019}, so by substituting $\mathcal{N}=2$ to the supplemental Eqs.~S32,~S47,~S66 of Ref.~\cite{Lemesh2018}, we can find
\begin{align}
\sigma_{d, s}^{1, 2 ({\text{SAF}})} &= \frac{2\pi \mu_{0}M_{s, 0}^2\Delta_0^2}{\mathcal{P}\mathcal{N}} G_s\left(\frac{\mathcal{T}}{2\pi \Delta_0}\right)+\frac{2\pi \mu_{0}M_{s, 1}^2\Delta_1^2}{\mathcal{P}\mathcal{N}} G_s\left(\frac{\mathcal{T}}{2\pi \Delta_1}\right)\notag \\ &+\text{sign}(J^{ex})\frac{2\mu_{0}M_{s, 0} M_{s, 1}}{\pi  \mathcal{N}\mathcal{P} } \int_{0}^{\infty}d k \left[\frac{\pi^2 }{4}\frac{\Delta_0 \Delta_{1}  k^2 }{\sinh\left(\frac{\pi  \Delta_0 k}{2}\right)\sinh\left(\frac{\pi  \Delta_{1} k}{2}\right)} -1\right] \frac{2e^{-k\mathcal{P}}-e^{-k(\mathcal{T}+\mathcal{P})}-e^{-k(\mathcal{T}-\mathcal{P})}}{k^3}\label{eq:appa:app:ss11}\\
\sigma_{d, v}^{1, 2 ({\text{SAF}})} &= \frac{2\pi \mu_{0}M_{s, 0}^2\Delta_0^2}{\mathcal{P}\mathcal{N}} G_v\left(\frac{\mathcal{T}}{2\pi \Delta_0}\right)\sin^2(\psi_0)+\frac{2\pi \mu_{0}M_{s, 1}^2\Delta_1^2}{\mathcal{P}\mathcal{N}} G_v\left(\frac{\mathcal{T}}{2\pi \Delta_1}\right)\sin^2(\psi_1)\notag \\ &+\frac{\pi \mu_{0}M_{s, 0} M_{s, 1}\Delta_0\Delta_{1}\sin(\psi_0)\sin(\psi_{1})}{2\mathcal{N}\mathcal{P}}  \int_{0}^{\infty}dk\frac{e^{-k(\mathcal{T}+P)}+e^{-k(\mathcal{T}-P)}-2 e^{-k\mathcal{P}}+2 k(\mathcal{T}-P)}{k\cosh\left(\frac{\pi \Delta_0 k}{2}\right)\cosh\left(\frac{\pi \Delta_{1} k}{2}\right)}	\label{eq:appa:app:vv11}\\
\sigma_{d, sv}^{1, 2 ({\text{SAF}})} 
& = \pm\frac{\pi \mu_{0}M_{s, 0} M_{s, 1} \Delta_0\Delta_1}{2\mathcal{N}\mathcal{P}}  \int_{0}^{\infty}dk\left[\frac{\text{sign}(J^{ex})\sin(\psi_1)}{\sinh\left(\frac{\pi \Delta_0 k}{2}\right)\cosh\left(\frac{\pi \Delta_1 k}{2}\right)}-\frac{\sin(\psi_0)}{\sinh\left(\frac{\pi \Delta_1 k}{2}\right)\cosh\left(\frac{\pi \Delta_0 k}{2}\right)} \right]\notag \\ & \times \frac{  e^{-k(P-\mathcal{T})}+e^{-k(P+\mathcal{T})}-2 e^{-k\mathcal{P}}}{k}\label{eq:appa:app:sv11}
\end{align}
Here, $J_{\text{ex}}$ is the interlayer exchange coupling constant, which is negative for the SAF structures. Equations~\eqref{eq:appa:app:ss11}-\eqref{eq:appa:app:sv11} express the exact magnetostatic energy of SAF layered structures. For simplicity, we can assume that inside the integral expressions, $\Delta_i=\Delta_j\equiv\Delta$ (i.e., equals to some average DW width), in which case they can be reduced to analytic expressions in the same manner as above and as in Ref.~\cite{Lemesh2018},
\begin{align}
\sigma_{d, s}^{1, 2 ({\text{SAF}})} &\approx \frac{2\pi \mu_{0}M_{s, 0}^2\Delta_0^2}{\mathcal{P}\mathcal{N}} G_s\left(\frac{\mathcal{T}}{2\pi \Delta_0}\right)+\frac{2\pi \mu_{0}M_{s, 1}^2\Delta_1^2}{\mathcal{P}\mathcal{N}} G_s\left(\frac{\mathcal{T}}{2\pi \Delta_1}\right)\notag \\ &+\text{sign}(J^{ex})\frac{2\pi \mu_{0}M_{s, 0}M_{s, 1}\Delta^2}{\mathcal{P}\mathcal{N}}\left[G_s\left(\frac{\mathcal{P}+\mathcal{T}}{2\pi\Delta}\right)+G_s\left(\frac{\mathcal{P}-\mathcal{T}}{2\pi\Delta}\right)-2G_s\left(\frac{\mathcal{P}}{2\pi\Delta}\right)\right]\label{eq:appa:app:ss111}\\
\sigma_{d, v}^{1, 2 ({\text{SAF}})} &\approx \frac{2\pi \mu_{0}M_{s, 0}^2\Delta_0^2}{\mathcal{P}\mathcal{N}} G_v\left(\frac{\mathcal{T}}{2\pi \Delta_0}\right)\sin^2(\psi_0)+\frac{2\pi \mu_{0}M_{s, 1}^2\Delta_1^2}{\mathcal{P}\mathcal{N}} G_v\left(\frac{\mathcal{T}}{2\pi \Delta_1}\right)\sin^2(\psi_1)\notag \\ &+\frac{2\pi \mu_{0}M_{s, 0}M_{s, 1}\Delta^2}{\mathcal{P}\mathcal{N}}\left[G_v\left(\frac{\mathcal{P}+\mathcal{T}}{2\pi\Delta}\right)+G_v\left(\frac{\mathcal{P}-\mathcal{T}}{2\pi\Delta}\right)-2G_v\left(\frac{\mathcal{P}}{2\pi\Delta}\right)\right]\sin(\psi_0)\sin(\psi_1)\label{eq:appa:app:vv111}\\
\sigma_{d, sv}^{1, 2 ({\text{SAF}})} 
& \approx \pm \frac{\pi \mu_{0}M_{s, 0}M_{s, 1}\Delta^2}{\mathcal{P}\mathcal{N}}\left[G_{sv}\left(\frac{\mathcal{P}+\mathcal{T}}{2\pi\Delta}\right)+G_{sv}\left(\frac{\mathcal{P}-\mathcal{T}}{2\pi\Delta}\right)-2G_{sv}\left(\frac{\mathcal{P}}{2\pi\Delta}\right)\right][\text{sign}(J^{ex})\sin(\psi_1)-\sin(\psi_0)]\label{eq:appa:app:sv111}
\end{align}
which, by using the introduced earlier $F_{\alpha, ij}=F_{\alpha, ij}(M_s, \Delta, \mathcal{T}, \mathcal{P}, \mathcal{N})$ convention, can be regrouped into $\sigma_{d}^{1, 2 ({\text{SAF}})}  = \sigma_{d, \text{intra}}^{1, 2 ({\text{SAF}})}+\sigma_{d, \text{inter}}^{1, 2 ({\text{SAF}})} $,
\begin{align}
\sigma_{d, \text{intra}}^{1, 2 ({\text{SAF}})} &= F_{s, 00}(M_{s, 0},\Delta_0) +F_{s, 00}(M_{s, 1},\Delta_1) +F_{v, 00}(M_{s, 0},\Delta_0)\sin^2(\psi_0) +F_{v, 00}(M_{s, 1},\Delta_1)\sin^2(\psi_1) \label{eq:appa:app:intra}\\
\sigma_{d, \text{inter}}^{1, 2 ({\text{SAF}})} &\approx  2\ \text{sign}(J^{ex})F_{s, 01}(\sqrt{M_{s, 0} M_{s, 1}},\Delta) +2\ F_{v, 01}(\sqrt{M_{s, 0} M_{s, 1}},\Delta)\sin(\psi_0) \sin(\psi_1) 	\notag \\&\pm F_{sv, 01}(\sqrt{M_{s, 0} M_{s, 1}},\Delta)[\text{sign}(J^{ex})\sin(\psi_1)-\sin(\psi_0)]\label{eq:appa:app:inter}
\end{align}

The intralayer interactions described by Eq.~\eqref{eq:appa:app:intra} clearly correspond to Eq.~\eqref{app:yang} originally used in Ref.~\cite{Yang2019} (although our expressions are more accurate). However, our derived equations also see show that this treatment omits the interlayer terms described by Eq.~\eqref{eq:appa:app:inter}. Such terms also contribute to the total equation of motion (Eqs.~15(a-c) in Ref.~\cite{Yang2019}) and their effect can be calculated by evaluating the corresponding $\partial \mathcal{L}/\partial \psi_i$ components of Eq.~\eqref{eq:psid0},
\begin{align}
\partial \sigma_{d, \text{inter}}^{1, 2 ({\text{SAF}})}/\partial \psi_0 &= \cos(\psi_0) [2\ F_{v, 01}(\sqrt{M_{s, 0} M_{s, 1}},\Delta)\sin(\psi_1) \mp F_{sv, 01}(\sqrt{M_{s, 0} M_{s, 1}},\Delta)]\label{eq:appa:app:intrader0}\\
\partial \sigma_{d, \text{inter}}^{1, 2 ({\text{SAF}})}/\partial \psi_1  &= \cos(\psi_1) [2\ F_{v, 01}(\sqrt{M_{s, 0} M_{s, 1}},\Delta)\sin(\psi_0)  \pm F_{sv, 01}(\sqrt{M_{s, 0} M_{s, 1}},\Delta)\text{sign}(J^{ex})]\label{eq:appa:app:interder1}
\end{align}
Going back to the variables and angle definitions used in Ref.~\cite{Yang2019}, we can renormalize these per unit length as
\begin{align}
\partial \lambda_{d, \text{inter}}^{1, 2 ({\text{SAF}})}/\partial \psi_L &= -\sin \psi_L \left[2\ F_{v, 01}\left(\sqrt{\frac{M_U M_L}{t_U t_L}},\Delta\right)\cos(\psi_U)  \mp F_{sv, 01}\left(\sqrt{\frac{M_U M_L}{t_U t_L}},\Delta\right)\right]\mathcal{P}\mathcal{N}\label{eq:appa:app:intrader00}\\
\partial \lambda_{d, \text{inter}}^{1, 2 ({\text{SAF}})}/\partial \psi_U  &= -\sin \psi_U \left[2\ F_{v, 01}\left(\sqrt{\frac{M_U M_L}{t_U t_L}},\Delta\right)\cos(\psi_L)  \pm \text{sign}(J^{ex}) F_{sv, 01}\left(\sqrt{\frac{M_U M_L}{t_U t_L}},\Delta\right)\right]\mathcal{P}\mathcal{N}\label{eq:appa:app:interder11},
\end{align}
These derived components contribute to the derivatives of the total energy, which from Eqs.~S16(a) and S16(b) of Ref.~\cite{Yang2019} are
\begin{align}
\frac{\partial \lambda}{\partial \psi_L}&=2M_L\Delta_L \left(-\frac{H^k_L}{2} \sin 2 \psi_L+\frac{\pi}{2} H_x \sin \psi_L-\frac{\pi}{2} H_y \cos \psi_L +\frac{\pi}{2} H_L^{\text{DM}}\sin \psi_L\right) -2\Delta_U \xi J^{ex} \sin(\psi_L-\psi_U)+ \frac{\partial \lambda_{d, \text{inter}}^{1, 2 ({\text{SAF}})}}{\partial \psi_L} \label{eq:app5:modified1}\\
\frac{\partial \lambda}{\partial \psi_U}&=2M_U\Delta_U \left(-\frac{H^k_U}{2}\sin 2\psi_U+\frac{\pi}{2}H_x\sin \psi_U-\frac{\pi}{2}H_y\cos \psi_U+\frac{\pi}{2}H^{\text{DM}}_U\sin \psi_U\right) -2\Delta_U\xi J^{ex} \sin(\psi_L-\psi_U)+ \frac{\partial \lambda_{d, \text{inter}}^{1, 2 ({\text{SAF}})}}{\partial \psi_U} \label{eq:app5:modified2}
\end{align}
and hence, the resulting DW equation of motion (provided by Eqs.~S15(a-c) of Ref.~\cite{Yang2019}) is also affected. 

If we insert the parameters from Ref.~\cite{Yang2019}, $M_L=\SI{6.3e5}{A/m \cdot nm}$, $M_U=\SI{3.9e5}{A/m \cdot nm}$, $t_u=t_l=\SI{1}{nm}$, $\Delta_U\approx \Delta_L=\SI{3.7}{nm}$, $J_{ex}=\SI{-0.15}{mJ/m^2}$, $\xi \approx2$, we can see that the exchange coefficient is $2\Delta_U\xi J^{ex}=\SI{2.2e-12}{J m^{-1} rad^{-1}}$, This value, assuming $\mathcal{P}=\SI{2}{nm}$, $\mathcal{N}=2$, significantly exceeds both the volume-volume component, which is $2\mathcal{P}\mathcal{N}F_{v, 01}\left(\sqrt{M_U M_L/(t_U t_L)},\Delta\right)\approx \SI{9.4e-14}{J m^{-1} rad^{-1}}$ and the surface-volume component, which is $\mathcal{P}\mathcal{N}F_{sv, 01}\left(\sqrt{M_U M_L/(t_U t_L)},\Delta\right)\approx\SI{1.9e-13}{J m^{-1} rad^{-1}}$. Thus, examining Eqs.~S15(a-c) of Ref.~\cite{Yang2019} and Eqs.~\eqref{eq:app5:modified1},~\eqref{eq:app5:modified2}, one can see that in AFM structures, the effect of precession arises due to the significant values of interlayer exchange coefficients and values of field $H_x$ (which result in irresolvable steady state equations). This is in sharp contrast with our magnetostatically coupled $\mathcal{N}\geq 3$ case, for which the precession is caused by a large number of relatively small magnetostatic terms (but also resulting in irresolvable steady state equations).

For this reason, our expectation is that for the synthetic bilayer AFM structures, magnetostatic effects contribute to the resulting precession only as a first order correction. The main driving force is the phenomenon of interlayer exchange coupling, magnetic moment imbalance, and negative bias field, in accordance with the findings of Ref.~\cite{Yang2019}.
\end{document}